\input harvmac
%\draftmode
\noblackbox %%%%%%%%%%%%%%%%% Lineskip
%%%%%%%%%%%%%%%%%%%%%%%%%%%%%%%%%%%%%%
\ifx\answ\bigans
\magnification=1200\baselineskip=14pt plus 2pt minus 1pt
\else\baselineskip=16pt % plus 2pt minus 1pt % 32 lines in l-format
\fi

%%%%%%%%%%%% Local definitons %%%%%%%%%%%%%%%%%%%%%%%%%%%%%%%%%%%%

\def\sss#1{{\scriptstyle #1}}
\def\INT{{\int\limits_0^1\hskip-0.1cm dx\hskip-0.15cm\int\limits_0^1
\hskip-0.1cm dy\hskip-0.15cm\int\limits_0^1 \hskip-0.1cm dz}}
\def\INTT{{\int\limits_0^1\hskip-0.1cm dx\hskip-0.15cm\int\limits_0^1
\hskip-0.1cm dy\hskip-0.15cm\int\limits_0^1 \hskip-0.1cm dz
\hskip-0.15cm\int\limits_0^1 \hskip-0.1cm dw}}

\def\ap{\alpha'}

\def\cf{{\it cf.\ }}
\def\ie{{\it i.e.\ }}

\def\eps{\epsilon}
\def\al{\alpha}

\def\si{\sigma}

\def\bet{\beta}\def\be{\beta}

\def\Si{{\Sigma}}

\def\comment#1{{}}

\def\FF#1#2{{_#1F_#2}}
%%%%%%%%%%%%%%%%%%%%%%%%%%%%%%%%%%%%%%%%%%%%%%%%%%%%%%%%%%%%%%%%
%%%%%   Dirac-Slash
%%%%%%%%%%%%%%%%%%%%%%%%%%%%%%%%%%%%%%%%%%%%%%%%%%%%%%%%%%%%%%%%
\def\slashchar#1{\setbox0=\hbox{$#1$}           % set a box for #1
   \dimen0=\wd0                                 % and get its size
   \setbox1=\hbox{/} \dimen1=\wd1               % get size of /
   \ifdim\dimen0>\dimen1                        % #1 is bigger
      \rlap{\hbox to \dimen0{\hfil/\hfil}}      % so center / in box
      #1                                        % and print #1
   \else                                        % / is bigger
      \rlap{\hbox to \dimen1{\hfil$#1$\hfil}}   % so center #1
      /                                         % and print /
   \fi}
%%%%%%%%%%%%%%%%%%%%%%%%%%%%%%%%%%%%%%%%%%%%%%%%%%%%%%%%%%%%%%%%%
%%%%% Referencing  %%%%%%%%%%%%%%%%%%%%%%%%%%%%%%%%%%%%%%%%%%%%%%
%%%%%%%%%%%%%%%%%%%%%%%%%%%%%%%%%%%%%%%%%%%%%%%%%%%%%%%%%%%%%%%%%
\newif\ifnref
\def\rrr#1#2{\relax\ifnref\nref#1{#2}\else\ref#1{#2}\fi}
\def\ldf#1#2{\begingroup\obeylines
\gdef#1{\rrr{#1}{#2}}\endgroup\unskip}

\def\doubref#1#2{\refs{{#1},{#2} }}
\def\threeref#1#2#3{\refs{{#1},{#2},{#3} }}

\nreffalse

\def\lref{\ldf}
%%%%%%%%%%%%%%%%%%%%%%%%%%%%%%%%%%%%%%%%%%%%%%%%%%%%%%%%%%%%%%%%%%
%%%%%%%%%%%%%%%%%   Stuff for Figures  %%%%%%%%%%%%%%%%%%%%%%%%%%%
%%%%%%%%%%%%%%%%%%%%%%%%%%%%%%%%%%%%%%%%%%%%%%%%%%%%%%%%%%%%%%%%%%

\input epsf
\input psfig
\def\figin{\epsfcheck\figin}\def\figins{\epsfcheck\figins}
\def\epsfcheck{\ifx\epsfbox\UnDeFiNeD
\message{(NO epsf.tex, FIGURES WILL BE IGNORED)}
\gdef\figin##1{\vskip2in}\gdef\figins##1{\hskip.5in}% blank space instead
\else\message{(FIGURES WILL BE INCLUDED)}%
\gdef\figin##1{##1}\gdef\figins##1{##1}\fi}
\def\DefWarn#1{}
\def\figinsert{\goodbreak\midinsert}
\def\ifig#1#2#3{\DefWarn#1\xdef#1{fig.~\the\figno}
\writedef{#1\leftbracket fig.\noexpand~\the\figno}%
\figinsert\figin{\centerline{#3}}\medskip\centerline{\vbox{\baselineskip12pt
\advance\hsize by -1truein\noindent\footnotefont{\bf Fig.~\the\figno } #2}}
\bigskip\endinsert\global\advance\figno by1}

%%%%%%%%%%%%%%%%%%%%%%%%%%%%%%%%%%%%%%%%%%%%%%%%%%%%%%%%%%%%%%%%%%%%%
%%%%%%%%%%%%%%%   Standard alltime definitions   %%%%%%%%%%%%%%%%%%%%
%%%%%%%%%%%%%%%%%%%%%%%%%%%%%%%%%%%%%%%%%%%%%%%%%%%%%%%%%%%%%%%%%%%%%

\def\appA{A}
\def\appB{B}
\def\appC{C}

\def\tilde{\widetilde}

\def\h {{1\over 2}}

\def\ov {\overline}
\def\o {\over}
\def\fc#1#2{{#1 \o #2}}

\def\IR{ {\bf R}}

      % For Eisenstein E2
  % For Polylogarithm

\def\br{\hfill\break}

\def\det {{\rm det}}

\def\lf {\left}
\def\ri {\right}

\def\p {\partial}

  \def\Nc{{\cal N}}\def\Wc{{\cal W}}
 \def\Qc{{\cal Q}}
 
 \def\Oc {{\cal O}}

\def\Ic {{\cal I}} 
\def\Kc {{\cal K}} 
  \def\Kc{{\cal K}}
%%%%%%%%%%%%%%%%%%%%%%%%%%%%%%%%%%%%%%%%%%%%%%%%%%%%%%%%%%%%%%%%%%%
%%%%%%%%%%%%%%%%%%%%%%%%%%%%%%%%%%%%%%%%%%%%%%%%%%%%%%%%%%%%%%%%%%%

\lref\KLLSW{
  V.A.~Kostelecky, O.~Lechtenfeld, W.~Lerche, S.~Samuel and S.~Watamura,
``Conformal Techniques, Bosonization and Tree Level String Amplitudes,''
  Nucl.\ Phys.\  B {\bf 288}, 173 (1987).
  %%CITATION = NUPHA,B288,173;%%
}

\lref\HK{A.~Hashimoto and I.R.~Klebanov,
``Scattering of strings from D-branes,''
  Nucl.\ Phys.\ Proc.\ Suppl.\  {\bf 55B}, 118 (1997)
  [arXiv:hep-th/9611214].
  %%CITATION = NUPHZ,55B,118;%%
}
\lref\AntoniadisEW{
  I.~Antoniadis,
  ``A Possible new dimension at a few TeV,''
  Phys.\ Lett.\  B {\bf 246}, 377 (1990).
  %%CITATION = PHLTA,B246,377;%%
}
\lref\ptmhv{
  S.J.~Parke and T.R.~Taylor,
``An Amplitude for $n$ Gluon Scattering,''
  Phys.\ Rev.\ Lett.\  {\bf 56}, 2459 (1986).
  %%CITATION = PRLTA,56,2459;%%
}

\lref\ptsusy{
  S.J.~Parke and T.R.~Taylor,
``Perturbative QCD Utilizing Extended Supersymmetry,''
  Phys.\ Lett.\  B {\bf 157}, 81 (1985)
  [Erratum-ibid.\  {\bf 174B}, 465 (1986)].
  %%CITATION = PHLTA,B157,81;%%
}

\lref\Berkovits{
  N.~Berkovits,
``An alternative string theory in twistor space for N = 4
super-Yang-Mills,''
  Phys.\ Rev.\ Lett.\  {\bf 93}, 011601 (2004)
  [arXiv:hep-th/0402045].
  %%CITATION = PRLTA,93,011601;%%
}

\lref\Anton{I.~Antoniadis, N.~Arkani-Hamed, S.~Dimopoulos and G.R.~Dvali,
``New dimensions at a millimeter to a Fermi and superstrings at a TeV,''
  Phys.\ Lett.\  B {\bf 436}, 257 (1998)
  [arXiv:hep-ph/9804398].
  %%CITATION = PHLTA,B436,257;%%
}
\lref\Lykken{J.D.~Lykken,
``Weak Scale Superstrings,''
  Phys.\ Rev.\  D {\bf 54}, 3693 (1996)
  [arXiv:hep-th/9603133].
  %%CITATION = PHRVA,D54,3693;%%
}
\lref\ManganoBY{
  M.~L.~Mangano and S.~J.~Parke,
  ``Multiparton amplitudes in gauge theories,''
  Phys.\ Rept.\  {\bf 200}, 301 (1991)
  [arXiv:hep-th/0509223].
  %%CITATION = PRPLC,200,301;%%
}
\lref\DixonWI{
  L.~J.~Dixon,
  ``Calculating scattering amplitudes efficiently,''
  arXiv:hep-ph/9601359.
  %%CITATION = HEP-PH/9601359;%%
}
\lref\Grisaru{
  M.~T.~Grisaru and H.~N.~Pendleton,
  ``Some Properties Of Scattering Amplitudes In Supersymmetric Theories,''
  Nucl.\ Phys.\  B {\bf 124}, 81 (1977).}
\lref\GrisaruB{
  M.~T.~Grisaru, H.~N.~Pendleton and P.~van Nieuwenhuizen,
  ``Supergravity And The S Matrix,''
  Phys.\ Rev.\  D {\bf 15}, 996 (1977).
  %%CITATION = PHRVA,D15,996;%%
}
\lref\BerendsME{
  F.~A.~Berends and W.~T.~Giele,
  ``Recursive Calculations for Processes with n Gluons,''
  Nucl.\ Phys.\  B {\bf 306}, 759 (1988).
  %%CITATION = NUPHA,B306,759;%%
}
\lref\Peskin{S.~Cullen, M.~Perelstein and M.E.~Peskin,
``TeV strings and collider probes of large extra dimensions,''
  Phys.\ Rev.\  D {\bf 62}, 055012 (2000)
  [arXiv:hep-ph/0001166].
  %%CITATION = PHRVA,D62,055012;%%
}

\lref\banksi{T.~Banks, L.J.~Dixon, D.~Friedan and E.J.~Martinec,
``Phenomenology and Conformal Field Theory Or Can String Theory Predict the
Weak Mixing Angle?,''
  Nucl.\ Phys.\  B {\bf 299}, 613 (1988)
  %%CITATION = NUPHA,B299,613;%%
}
\lref\banksii{T.~Banks and L.J.~Dixon,
``Constraints on String Vacua with Space-Time Supersymmetry,''
  Nucl.\ Phys.\  B {\bf 307}, 93 (1988).
  %%CITATION = NUPHA,B307,93;%%
}

\lref\JOE{
J. Polchinski, "String Theory'', Sections 6 \& 12, Cambridge University Press 1998.}

\lref\flt{
  S.~Ferrara, D.~L\"ust and S.~Theisen,
 ``World Sheet Versus Spectrum Symmetries In Heterotic And Type II Superstrings,''
  Nucl.\ Phys.\  B {\bf 325}, 501 (1989).
  %%CITATION = NUPHA,B325,501;%%
}
\lref\PCJ{
  J.~Polchinski, S.~Chaudhuri and C.V.~Johnson,
``Notes on D-Branes,''
  arXiv:hep-th/9602052.
  %%CITATION = HEP-TH/9602052;%%
}
\lref\Brazili{
R.~Medina and L.A.~Barreiro,
``Higher N-point amplitudes in open superstring theory,''
  arXiv:hep-th/0611349.
  %%CITATION = HEP-TH/0611349;%%
}

\lref\Brazil{L.A.~Barreiro and R.~Medina,
``5-field terms in the open superstring effective action,''
  JHEP {\bf 0503}, 055 (2005)
  [arXiv:hep-th/0503182];\br
  %%CITATION = JHEPA,0503,055;%%
R.~Medina, F.T.~Brandt and F.R.~Machado,
``The open superstring 5-point amplitude revisited,''
  JHEP {\bf 0207}, 071 (2002)
  [arXiv:hep-th/0208121].
  %%CITATION = JHEPA,0207,071;%%
}

\lref\FMS{
  D.~Friedan, E.J.~Martinec and S.H.~Shenker,
``Conformal Invariance, Supersymmetry And String Theory,''
  Nucl.\ Phys.\  B {\bf 271}, 93 (1986).
  %%CITATION = NUPHA,B271,93;%%
}

\lref\Dan{D.~Oprisa and S.~Stieberger,
``Six gluon open superstring disk amplitude, multiple hypergeometric  series
and Euler-Zagier sums,''
  arXiv:hep-th/0509042.
  %%CITATION = HEP-TH 0509042;%%
}

\lref\Witten{
  E.~Witten,
 ``Perturbative gauge theory as a string theory in twistor space,''
  Commun.\ Math.\ Phys.\  {\bf 252}, 189 (2004)
  [arXiv:hep-th/0312171].
  %%CITATION = CMPHA,252,189;%%
}

\lref\CSW{F.~Cachazo, P.~Svrcek and E.~Witten,
``MHV vertices and tree amplitudes in gauge theory,''
  JHEP {\bf 0409}, 006 (2004)
  [arXiv:hep-th/0403047].
  %%CITATION = JHEPA,0409,006;%%
}

\lref\STii{
  S.~Stieberger and T.R.~Taylor,
``Multi-gluon scattering in open superstring theory,''
  Phys.\ Rev.\  D {\bf 74}, 126007 (2006)
  [arXiv:hep-th/0609175].
  %%CITATION = PHRVA,D74,126007;%%
}

\lref\STi{
  S.~Stieberger and T.R.~Taylor,
``Amplitude for N-gluon superstring scattering,''
  Phys.\ Rev.\ Lett.\  {\bf 97}, 211601 (2006)
  [arXiv:hep-th/0607184].
  %%CITATION = PRLTA,97,211601;%%
}

\lref\STiii{
S. Stieberger and T.R.~Taylor, work in progress.}

%%%%%%%%%%%%%%%%%%%%%%%%%%%%%%%%%%%%%%%%%%%%%%%%%%%%%%%%%%%%%%%%%%%%%%%%%%%%%%%
\Title{\vbox{\rightline{MPP-2007-79}\rightline{LMU--ASC 51/07}
}}
{\vbox{\centerline{Supersymmetry Relations and MHV Amplitudes}
\bigskip\centerline{in Superstring Theory}
}}
\smallskip
\centerline{Stephan Stieberger$^{a}$\ \ and\ \ Tomasz R. Taylor$^{a,b,c}$}
\bigskip
\centerline{\it $^a$ Max--Planck--Institut f\"ur Physik}
\centerline{\it Werner--Heisenberg--Institut}
\centerline{\it 80805 M\"unchen, Germany}
\vskip7pt
\centerline{\it $^b$ Arnold--Sommerfeld--Center for Theoretical Physics}
\centerline{\it Ludwig--Maximilians--Universit\"at M\"unchen}
\centerline{\it 80333 M\"unchen, Germany}
\vskip7pt
\centerline{\it $^c$ Department of Physics}
\centerline{\it Northeastern University}
\centerline{\it Boston, MA 02115, USA}

% \medskip
\bigskip\bigskip
\centerline{\bf Abstract}
\vskip7pt
\noindent
We discuss supersymmetric Ward identities relating various scattering amplitudes
in type I open superstring theory. We show that at the disk level, the form of
such relations remains exactly the same, to all orders in $\alpha'$, as in the
low--energy effective field theory describing the $\alpha'\to 0$ limit. This
result holds in $D=4$ for all compactifications, even for those that break
supersymmetry. We apply SUSY relations to the computations of $N$--gluon MHV
superstring amplitudes, simplifying the existing results for $N\le 6$ and deriving
a compact expression for $N=7$.

\Date{}
\noindent

\goodbreak

%%%%%%%%%%%%%%%%%%%%%%%%%%%%%%%%%%%%%%%%%%%%%%%%%%%%%%%%%%%%%%%%%%%%%%%%%%%%%%%
\newsec{Introduction}

For the last thirty years, multi-gluon amplitudes and their supersymmetric
variants
have been extensively studied in the framework of quantum relativistic
theory of gauge fields. These amplitudes are very important from both
theoretical and experimental points of view because they describe the
scattering processes underlying hadronic jet production at high energy
colliders. The modern tools for amplitude computations\foot{For reviews, see \ManganoBY\ and \DixonWI.} include helicity
techniques, string-inspired color ordering and the use of supersymmetry
\refs{\ptsusy,\Grisaru,\GrisaruB} and recursion relations \BerendsME. More recently, a substantial progress has
been achieved by employing some elements of twistor theory \Witten.
In particular, a new type of recursion relations has been constructed,
inspired by twistor string theory \CSW. Here, similarly to the case of color
ordering, string theory comes as a handy tool.

Multi-gluon scattering is also a process of considerable theoretical
interest in the framework of full-fledged superstring theory, and would
have a great experimental relevance if the string mass scale turned out to
be low enough to be reached at LHC \doubref\Lykken\Anton.
In this case, some effects due to the off- and on-shell propagation of the
string Regge excitations could be detected in jet cross sections. The
parameter that controls the size of such effects is the Regge slope
$\alpha'$. As pointed out in \doubref\Lykken\Anton, a consistent low string
mass (small $\alpha'$) scenario must necessarily involve large extra
dimensions \AntoniadisEW, therefore massive string excitations come together with
Kaluza-Klein states.
However, the effects of Regge states dominate over the effects of extra
dimensions
for a wide range of parameters, notably in the weak string coupling regime
\Peskin.

In two recent papers \doubref\STi\STii, we initiated a systematic study of
multi-gluon scattering processes in open superstring theory, at the
semiclassical, disk level of the string worldsheet. We focussed
on one particular, maximum helicity violating (MHV) gluon helicity
configuration, because in the $\alpha'\to 0$ limit, the respective
amplitude is described by the well-known, simple formula \ptmhv\ 
(written in the notation of \doubref\ManganoBY\DixonWI):
\eqn\mhvform{A_{Y\! M}(1^-,2^-,3^+,4^+,\dots, N^+) = (\sqrt{2}\,g_{Y\! M})^{N-2}\,
{\rm Tr}  (\, T^{a_1}\cdots T^{a_N})\,{\cal M}_{Y\!M}^{(N)}
}
where $g_{Y\! M}$ is the properly normalized Yang-Mills coupling constant and
\eqn\mhvpt{{\cal M}_{Y\!M}^{(N)}= i\
{\langle 12\rangle^4\over\langle 12\rangle\langle 23\rangle\langle
34\rangle\cdots\langle N1\rangle .}
}
{}For $N\le 6$
external gluons, we presented the amplitudes in a factorized form, as a
product of the zero slope (Yang-Mills) MHV amplitude ${\cal M}_{Y\!M}^{(N)}$
times a string ``formfactor''
involving $(N{-}3)!$ generalized hypergeometric functions of kinematic
invariants.
We argued that the
soft and collinear factorization properties, combined with the Abelian
limit, are completely sufficient to determine all $N$-gluon MHV
amplitudes. Nevertheless, the increasing complexity of string formfactors
has not yet allowed us writing a compact expression describing  MHV string
amplitudes with arbitrary numbers of gluons.

In this work, we show that the computations of $N$-gluon string amplitudes
can be simplified by utilizing supersymmetric Ward identities, exactly in
the same way as in $\alpha'=0$ gauge theories \ptsusy. In Section 1, we
prove that at the disk level, the well-known supersymmetry (SUSY) relations
\ptsusy\ between amplitudes involving gluons, gluinos and gauge scalars are
valid to all orders in $\alpha'$. Since the gluonic disk amplitudes are
completely determined by the four-dimensional ($D=4$) spacetime part of the
underlying world-sheet superconformal field theory (SCFT), these relations
can be used to simplify them even if supersymmetry is broken by the
compactification.
In particular, for any compactification of open superstring theory, the
$N$-gluon MHV disk amplitude can be expressed in terms of an amplitude
involving $N{-}4$ gluons and 4 scalars. In Section 2, we use this relation
to rewrite the results of \doubref\STi\STii\ in a non-factorized way, but
in terms of  ``primitive'' (hypergeometric) integrals. This non-factorized
form of amplitudes looks more promising for constructing a string
generalization of
MHV rules \CSW\ and it could potentially lead to some soluble recursion
relations \STiii.
In Section 3, we demonstrate the power of SUSY relations by deriving a
compact expression for the seven-gluon MHV string amplitude.
In Section 4 we compute various four--point amplitudes involving scalars, gauginos and
vectors and present an explicit example of a SUSY relation.

\newsec{Spacetime supersymmetry relations and string amplitudes}

In superstring theory, the vertices creating gauge bosons, gauginos and
gauge scalars are related by supersymmetry transformations. Here,
we use the operator product expansion (OPE) rules in order to
evaluate the corresponding contour integrals, with SUSY transformations
generated by the appropriate insertions of (spacetime) supercharge
operators. We use these transformations in order to derive
supersymmetric Ward identities for some amplitudes related to $N$-gluon scattering.

This Section is divided into four parts. In the first part, we use the so-called ``doubling trick'' \HK\ to
rewrite the commutators of (spacetime) SUSY generators with open string vertices, inserted at the boundary of a disk
world-sheet, as the contour integrals on the sphere.
In the second part, we apply these contour integrals
in order to determine the SUSY commutators
for the vertices creating full gauge supermultiplet.
In the third part, we explain how to use contour deformations in
order to derive supersymmetric Ward identities relating various superstring
scattering amplitudes. Finally, in the last part
we  construct the helicity basis for string vertex operators, and show that their SUSY transformations agree with the known results in the $\alpha'\to 0$
(Yang-Mills) limit.

\subsec{Extended spacetime supersymmetry and SUSY variations on the disk}

In the following, we establish SUSY transformation
rules on the open string disk world-sheet.
We consider type II or type I superstring
compactifications on an internal six-dimensional manifold with
$2\cal N$ or $\cal N$ extended supersymmetry charges in $D=4$ spacetime, respectively.
In type II superstring theory, one
can construct,  on a closed string world-sheet, two  sets of $\cal N$ holomorphic
charges $\Qc^I_\al$ and $\tilde \Qc^I_\al$, $I=1,\dots,\cal N$,
with independent actions on the left- and right-moving closed string
modes, respectively.
The supercharges $\Qc^I_\al$ of the left-moving sector are given by
the contour integrals (fixed time lines on the closed string world-sheet)
\eqn\susy{
\Qc_\al^I=\oint \fc{dz}{2\pi i}\ V_\al^I (z)\ \ \ ,\ \ \
\ov \Qc^I_{\dot \al}=\oint \fc{dz}{2\pi i}\ \ov V^I_{\dot \al} (z)\ ,}
of the holomorphic supercurrents $V^I_\al (z)$ and
$\ov V^I_{\dot \al} (z)$.
In $D=4$, in the $(-1/2)$-ghost picture \FMS, these are given by
\eqn\Susycurrents{
V^I_\al(z)=\ap^{-1/4}\ e^{-\h\phi(z)}\ S_\al(z)\ \Si^I(z)\ \ ,\ \ \
\ov V^I_{\dot\al}(z)=\ap^{-1/4}\ e^{-\h\phi(z)}\ S_{\dot\al}(z)\ \ov\Si^I(z)\ ,}
where $S_\al,S_{\dot\al}$ are the spin fields with the
indices $\al$ (or $\dot\al$)  denoting negative (positive) chirality
in four dimensions.  The
Ramond fields $\Si^I$ belong to an internal SCFT with $c=9$.
Finally, $\phi$ is the scalar bosonizing the superghost system.
The OPEs of the internal Ramond fields~$\Sigma^I$ are \banksii\ (\cf also \flt)
\eqn\ope{\eqalign{
\Sigma^I(z)\ \ov \Sigma^J(w)&=(z-w)^{-3/4}\ \delta^{IJ}+(z-w)^{1/4}\
J^{IJ}(w)+\ldots\ ,\cr
\Sigma^I(z)\ \Sigma^{J}(w)&=(z-w)^{-1/4}\ \psi^{IJ}(w)+\Oc((z-w)^{3/4})\ ,}}
with the dimension one currents $J^{IJ}$ and dimension $1/2$ operators $\psi^{IJ}$.
Additional material on OPEs is given in the Appendix \appA,
where we also summarize some basic facts about type II SUSY algebra.
The right-moving supercharges $\tilde \Qc^I_\al$ and
$\ov{\tilde \Qc}^I_\al$ are constructed in the same way from the
anti-holomorphic currents $\tilde {V}^I_\al(\bar z)$ and
$\ov{\tilde V}^I_{\dot\al}(\bar z)$, respectively.

In type I open superstring theory the left- and
right-movers are tied
together by the world--sheet boundaries and only the total charge
\eqn\only{
Q_\al^I=\Qc^I_a+\tilde \Qc_\al^I~,}
acting on the open string modes, is preserved. The combination \only\
enjoys then the open string SUSY algebra.
On an open string world-sheet, some boundary conditions have to be imposed on the
fields. On the disk, which is conformally
equivalent to the upper half plane ${\bf H}_+$, the following
boundary conditions are imposed on the supersymmetry currents:
\eqn\impose{
V_\al^I(z)=\tilde V_\al^I(\bar z)\ \ \ ,\ \ \ z=\bar z\ .}
It is convenient to extend the definition of the currents
$V_\al^I$ to the full complex plane by using the so-called doubling trick \HK:
\eqn\doubling{
V_\al^I(z)=\cases{ V_\al^I(z)\ \ \ ,&        $z\in {\bf H}_+\ ,$\cr
                 \tilde V_\al^I(z)\ \ \ ,& $z\in {\bf H}_-\ .$}}
On the disk ${\bf H}_+$ a fixed time integral is represented by a
half--circle around $z=0$ in the upper half plane.
% Integration over this ring of the full supercurrent $V_\al^I(z)+\tilde
% V_\al^I(\ov z)$ gives combines into a closed contour integral
% over the field \doubling\ in the full complex plane.
Now, the line (fixed time) integrals on the disk ${\bf H}_+$ can be combined
to full closed contour integrals on the sphere. In this way,
the conserved supersymmetry charges \only\ can be rewritten as
\eqn\Consider{
Q_\al^I=\int\fc{dz}{2\pi i}\ V_\al^I(z)+\int\fc{d\bar z}{2\pi i}\
\tilde V^I_\al(\bar z):=\oint \fc{dz}{2\pi i}\ V_\al^I(z)\ ,}
where on the l.h.s., the $z$ and $\ov z$ integrals are over semicircles in 
the upper and lower half-planes (${\bf H_+}$ and ${\bf H_-}$), respectively,
while the integral on the r.h.s. is over a circle on the full complex plane 
(Riemann sphere), see Fig. 1.
\ifig\figi{Fixed time lines $C_1$ and $C_2$ on the disk ${\bf
H_+}$ and their extensions to a closed loop $C$ on the sphere.}
{\epsfxsize=0.4\hsize\epsfbox{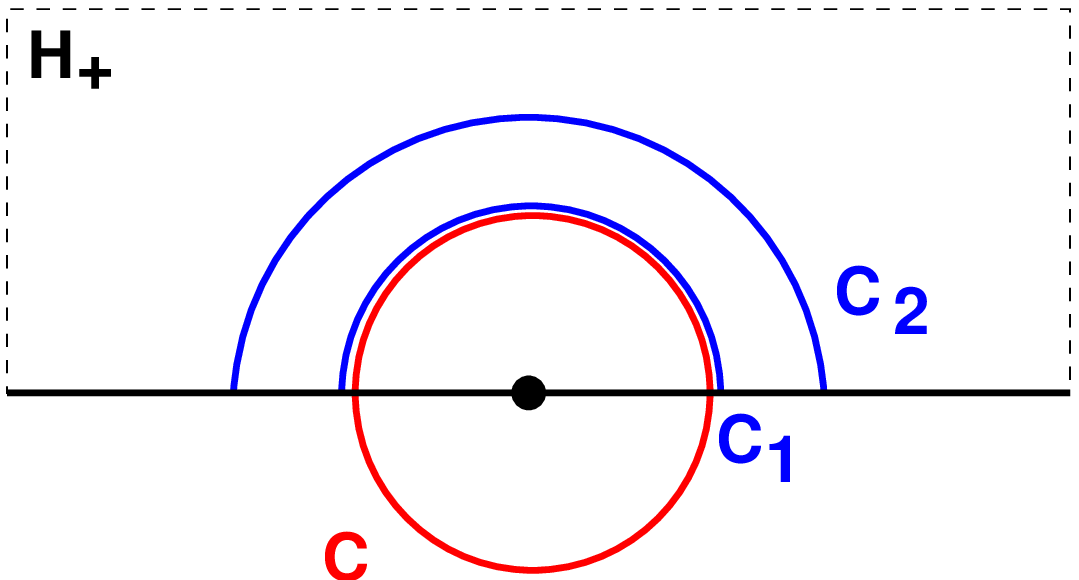}}
\noindent Hence when converting from type II to type I we can use the fields from the
left-moving closed string sector, extend their definition to the full complex
plane in lines of \doubling\ and consider contour integrals in the
full complex plane as in  Eq.\Consider. A similar procedure can be applied in
the presence of D$p$-branes that can further reduce the amount of conserved
supercharges, however for our purposes, $D=4$ type I compactifications are
completely sufficient.

In order to determine the variation of an arbitrary open string vertex
operator ${\cal O}(z)$ (inserted at the boundary $z=\bar z$)
under infinitesimal SUSY transformations generated by
\eqn\ssdef{Q^I(\eta_I,\bar\eta_I)=Q^I(\eta_I)+\bar Q^I(\bar\eta_I) \quad{\rm with}\quad
Q^I(\eta_I)=\eta_I^{\alpha}Q^I_{\alpha}~,\quad \bar
Q^I(\bar\eta_I)=\bar\eta_{I\dot\alpha}
\bar Q^{\dot\alpha I}}
one needs the commutators
\eqn\CONTOUR{
[Q^I(\eta_I),{\cal O}(z)]:=\oint\limits_{C_{\varepsilon}(z)}
\fc{dw}{2\pi i}\ \eta_I^{\al}\ V^I_{\al}(w)\ {\cal O}(z)\ ,}
where $C_{\varepsilon}(z)$ is a circle of radius $\varepsilon\to 0$
surrounding
the point $z$ and $V$ is the supercurrent extended
to the full complex plane according to Eq.\doubling. Thus the integral
\CONTOUR\ effectively picks up the residuum at $w=z$, hence it can be
determined by using OPE. One also needs similar commutators with $\bar Q^I(\bar\eta_I)$.

\subsec{SUSY transformations of open string vertices}
{}For our purposes, it is completely sufficient
to consider a maximally supersymmetric, toroidal compactification of
type I (or type IIA/B) superstring, with ${\cal N}=4$ SUSY in $D=4$.
Therefore in the following, we shall specialize to the this case, \ie $I,J=1,\ldots,4$.
{}For ${\cal N}=4$, the internal fields $\Si^I$ have an explicit realization as pure
exponentials \banksii:
\eqn\Banks{
\Si^1=e^{\fc{i}{2}(H_1+H_2+H_3)}\ ,\ \Si^2=e^{\fc{i}{2}(H_1-H_2-H_3)}\
,\ \Si^3=e^{\fc{i}{2}(-H_1+H_2-H_3)}\ ,\ \Si^4=e^{\fc{i}{2}(-H_1-H_2+H_3)}\ .}
Furthermore, the fields $\psi^{IJ}$ in \ope\ become complex fermions $\Psi^{i-}=e^{iH_i},\
\Psi^{i+}=e^{-iH_i}$:
\eqn\Dixon{
\psi^{12}=\ov\psi^{34}=e^{iH_1}\ \ ,\ \
\psi^{13}=\ov\psi^{24}=e^{iH_2}\ \ ,\ \
\psi^{14}=\ov\psi^{23}=e^{iH_3}\ .}
The fields $\psi^{IJ}$ are anti--symmetric w.r.t. to their internal indices $I$ and $J$.

The massless open string modes on the D$p$--brane world--volume
form an ${\cal N}=4$ gauge vector multiplet: three complex scalars $\phi^i$,
four gauginos $\lambda^I$ and one vector $A_\mu$ in the adjoint representation of the gauge group.
The scalar vertex, in the $(-1)$-ghost picture, reads
\eqn\fieldsi{\hskip -2cm
V_{\phi^{a,j\pm}}^{(-1)}(z,k) ~=~ g_\phi\ T^a\ e^{-\phi}\ \Psi^{j\pm}\
e^{ikX}\ \ \ \ ,\ \ \ j=1,\ldots,3\ ,}
with the complex fermions $\Psi^{j\pm}=
\fc{1}{\sqrt2}(\psi^{2j+2}\mp i\psi^{2j+3})$.
The gaugino vertex operators, in the $(-1/2)$-ghost picture, are
\eqn\fieldsii{\eqalign{
V_{\lambda^{a,I}}^{(-1/2)}(z,u,k)
&=g_\lambda\ T^a\ e^{-\phi/2}\ u^\al\ S_\al\ \Si^I\ e^{ik_\rho X^\rho}\ ,\cr
V_{\ov\lambda^{a,I}}^{(-1/2)}(z,u,k)
&=g_\lambda\ T^a\ e^{-\phi/2}\ \ov u_{\dot\bet}\ S^{\dot\bet}\ \ov\Si^I\ e^{ik_\rho X^\rho}
\ \ \ ,\ \ \ I=1,\ldots,4\ .}}
The gauge boson vertex operator in the $(-1)$-ghost picture reads:
\eqn\fieldsiii{\hskip -3.7cm
V_{A^a}^{(-1)}(z,\xi,k) ~=~ g_A\ T^a\ e^{-\phi}\ \xi^\mu\ \psi_\mu\ e^{ikX}\ .}
The open string vertex couplings are
\eqn\opscoupling{
g_\phi=(2\ap)^{1/2}\ g_{Y\! M}\ \ \ ,\ \ \ g_\lambda=(2\ap)^{1/2}\ap^{1/4}\ g_{Y\! M}\ \ \ ,\ \ \
g_A=(2\ap)^{1/2}\ g_{Y\! M}}
for the scalar, gaugino and vector, respectively \JOE.
The $D=4$ gauge
coupling $g_{Y\! M}$ can be expressed in terms of the ten--dimensional gauge coupling $g_{10}$
and the dilaton field $\phi_{10}$ through the relation $g_{Y\! M}=g_{10}e^{\phi_{10}/2}$ \JOE.
In the above definitions, $T^a$ are the Chan--Paton factors accounting for the gauge
degrees of freedom of the two open string ends.
Furthermore, the on--shell
constraints $k^2=0,\ \slashchar{k}u=0$ are imposed.

The vertices creating gauge bosons, gauginos and gauge
scalars are related by ${\cal N}=4$ SUSY transformations  \ssdef.
The sixteen Grassmann parameters  $\eta_I^\al,\bar\eta_{\dot\al I}$
are chiral spinors of both $SL(2,C)$ Lorentz group and the internal $SO(6)$ R-symmetry group.
Here, we use the OPE rules in order to evaluate the corresponding contour
integrals \CONTOUR. As an example,
$$\eqalign{
[\ \ov Q^I(\bar\eta_I) ,\ V_{\lambda^{a,J}}\ ]
&=g_A\ T^a\ \oint\limits_{w=z} \fc{dw}{2\pi i}\
e^{-\h\phi(w)}\ \ov\eta_{I\dot\al}S^{\dot\al}(w)\ \ov\Si^{I}(w)\
e^{-\h\phi(z)}\ u^\bet S_\bet(z)\ \Si^J(z)\ e^{ikX}\cr
&=\fc{g_A}{\sqrt2}\ T^a\ \oint\limits_{w=z} \fc{dw}{2\pi i}\
\fc{e^{-\phi(z)}}{(w-z)^{1/4}}\
(\ov\eta_{I\dot\al}\ov\sigma^{\mu\dot\al\bet}u_\bet)\ \fc{\psi_\mu(z)}{(z-w)^0}\
\fc{\delta^{IJ}}{(w-z)^{3/4}}\ e^{ikX}\cr
&=\fc{g_A}{\sqrt2}\ T^a\ e^{-\phi}\ \delta^{IJ}\
(\ov\eta_{I\dot\al}\ov\sigma^{\mu\dot\al\bet}u_\bet)
\ \psi_\mu\ e^{ikX}\ ,}$$
where we  used the OPEs given in \ope\ and in Appendix \appA. All remaining
commutators, obtained in a similar way, are collected below.

Applying the transformation \CONTOUR\ on the scalar vertex operators
\fieldsi, we obtain:
\eqn\susytransi{\eqalign{
[\ Q^I(\eta_I) ,\ V_{\phi^{a,i\pm}}\ ]&=g_\lambda\
T^a\ e^{-\h\phi}\ k_\mu\ (\eta^\al_I\
\sigma^\mu_{\al\dot{\beta}}\ S^{\dot\beta})\
\ov\Si^{J}\ e^{ikX}\ ,\cr
[\ \ov Q^I(\bar\eta_I) ,\ V_{\phi^{a,i\pm}}\ ]&=g_\lambda\ T^a\ e^{-\h\phi}\
k_\mu\ (\ov\eta_{I\dot\al}\
\ov\sigma^{\mu\dot\al\beta}S_{\beta})\ \Si^J\ e^{ikX}\ .}}
The respective internal indices $J$ for the Ramond fields (gaugino vertices) are given in
Tables 1a and 1b.
\vskip0.5cm\hskip-4.3cm
{\vbox{\ninepoint{
$$
\vbox{\offinterlineskip\tabskip=0pt
\halign{\strut\vrule#
%%%%%%%%%%%%%%%%%%
&~$#$~\hfil
&\vrule#
&~$#$~\hfil
&~$#$~\hfil
&~$#$~\hfil
&~$#$~\hfil
&~$#$~\hfil
&~$#$~\hfil
&\vrule#
&~$#$~\hfil
&~$#$~\hfil
&~$#$~\hfil
&~$#$~\hfil
&~$#$~\hfil
&~$#$~\hfil
&\vrule#&\vrule#
\cr
%%%%%%%%%%%%%%%%%%
\noalign{\hrule}
&
% [Q^I(\eta_I),V_{\star}]
&&&
\phi^{1-}
&&
\phi^{2-}
&&
\phi^{3-}
&&&
\phi^{1+}
&&
\phi^{2+}
&&
\phi^{3+}
&
\cr
%%%%%%%%%%%%%%%%%%
\noalign{\hrule}
&
Q^1
&&&
0
&&
0
&&
0
&&&
\ov\Si^{2}
&&
\ov\Si^{3}
&&
\ov\Si^{4}
&
\cr
%%%%%%%%%%%%%%%%%%
% \noalign{\hrule}
&
Q^2
&&&
0
&&
\ov\Si^{4}
&&
\ov\Si^{3}
&&&
\ov\Si^{1}
&&
0
&&
0
&
\cr
%%%%%%%%%%%%%%%%%%
% \noalign{\hrule}
&
Q^3
&&&
\ov\Si^{4}
&&
0
&&
\ov\Si^{2}
&&&
0
&&
\ov\Si^{1}
&&
0
&
\cr
%%%%%%%%%%%%%%%%%%
%\noalign{\hrule}
&
Q^4
&&&
\ov\Si^{3}
&&
\ov\Si^{2}
&&
0
&&&
0
&&
0
&&
\ov\Si^{1}
&
\cr
%%%%%%%%%%%%%%%%%%
%%%%%%%%%%%%%%%%%%
\noalign{\hrule}}}$$
\vskip-10pt
\centerline{\noindent{\bf Table 1a:}
{\sl Index structure $(I,i)$ and $J$ of the}}
\centerline{\sl supersymmetry variations $[ Q^I(\eta_I),V_{\phi^{i\pm}}]$.}
}}}
{\hskip-6.8cm\vbox{\ninepoint{
$$
\vbox{\offinterlineskip\tabskip=0pt
\halign{\strut\vrule#
%%%%%%%%%%%%%%%%%%
&~$#$~\hfil
&\vrule#
&~$#$~\hfil
&~$#$~\hfil
&~$#$~\hfil
&~$#$~\hfil
&~$#$~\hfil
&~$#$~\hfil
&\vrule#
&~$#$~\hfil
&~$#$~\hfil
&~$#$~\hfil
&~$#$~\hfil
&~$#$~\hfil
&~$#$~\hfil
&\vrule#&\vrule#
\cr
%%%%%%%%%%%%%%%%%%
\noalign{\hrule}
&
% [\ov Q^I(\bar\eta_I),V_\ast]
&&&
\phi^{1-}
&&
\phi^{2-}
&&
\phi^{3-}
&&&
\phi^{1+}
&&
\phi^{2+}
&&
\phi^{3+}
&
\cr
%%%%%%%%%%%%%%%%%%
\noalign{\hrule}
&
\ov Q^1
&&&
\Si^{2}
&&
\Si^{3}
&&
\Si^{4}
&&&
0
&&
0
&&
0
&
\cr
%%%%%%%%%%%%%%%%%%
% \noalign{\hrule}
&
\ov Q^2
&&&
\Si^{1}
&&
0
&&
0
&&&
0
&&
\Si^{4}
&&
\Si^{3}
&
\cr
%%%%%%%%%%%%%%%%%%
% \noalign{\hrule}
&
\ov Q^3
&&&
0
&&
\Si^{1}
&&
0
&&&
\Si^{4}
&&
0
&&
\Si^{2}
&
\cr
%%%%%%%%%%%%%%%%%%
%\noalign{\hrule}
&
\ov Q^4
&&&
0
&&
0
&&
\Si^{1}
&&&
\Si^{3}
&&
\Si^{2}
&&
0
&
\cr
%%%%%%%%%%%%%%%%%%
%%%%%%%%%%%%%%%%%%
\noalign{\hrule}}}$$
\vskip-10pt
\centerline{\noindent{\bf Table 1b:}
{\sl Index structure $(I,i)$ and $J$ of the}}
\centerline{\sl supersymmetry variations $[\ov Q^I(\bar\eta_I),V_{\phi^{i\pm}}]$.}
}}}
\vskip0.25cm
\hskip-0.75cm For gaugino vertices \fieldsii, we obtain
\eqn\susytransii{\eqalign{
[\ Q^I(\eta_I),V_{\lambda^{a,J}}\ ]&=g_\phi\ T^a\
e^{-\phi}\ (\eta_Iu)\ \psi^{IJ}\ e^{ikX}\ ,\cr
[\ \ov Q^I(\bar\eta_I),V_{\lambda^{a,J}}\ ]&=\fc{g_A}{\sqrt 2}\ T^a\
e^{-\phi}\ \delta^{IJ}\ (\ov\eta_{I\dot\al}\ov\sigma^{\mu\dot\al\beta}u_\beta)\
\psi_\mu\ e^{ikX}\ ,\cr
[\ Q^I(\eta_I),V_{\ov\lambda^{a,J}}\ ]&=\fc{g_A}{\sqrt 2}\ T^a\
e^{-\phi}\ \delta^{IJ}\ (\eta^\al_I\ \sigma_{\al\dot\beta}^\mu \ov u^{\dot\beta})\
\psi_\mu\ e^{ikX}\ ,\cr
[\ \ov Q^I(\bar\eta_I),V_{\ov\lambda^{a,J}}\ ]&=g_\phi\ T^a\
e^{-\phi}\ (\ov\eta_I \ov u)\ \ov \psi^{IJ}
\ e^{ikX}\ ,}}
with the complex fermions $\psi^{IJ}$ defined in \Dixon. Finally,
for the vectors \fieldsiii, we obtain
\eqn\susytransiii{\eqalign{
[\ Q^I(\eta_I) ,\ V_{A^a}\ ]&=\fc{g_\lambda}{\sqrt 2}\ T^a\ e^{-\h\phi}\ \xi_\mu\ k_\nu\
 \eta^\al_I\ (\sigma^{\mu\nu})_\al^{\ \bet}\ S_\bet\ \Si^{I}\ e^{ikX}\ ,\cr
[\ \ov Q^I(\bar\eta_I) ,\ V_{A^a}\ ]&=\fc{g_\lambda}{\sqrt 2}\
T^a\ e^{-\h\phi}\ \xi_\mu\ k_\nu\ \ov\eta_{I\dot\al}\
(\ov\sigma^{\mu\nu})^{\dot\al}_{\ \dot\bet}\ S^{\dot\bet}\ \ov\Si^{I}\ e^{ikX}\ ,}}
with
$(\sigma^{\mu\nu})_{\al}^{\ \bet}=\h(\si_{\al\dot\al}^\mu\ov\si^{\nu\dot\al\bet}-
\si_{\al\dot\al}^\nu\ov\si^{\mu\dot\al\bet})$ and
$(\ov\sigma^{\mu\nu})^{\dot\al}_{\ \dot\bet}=\h(\ov\si^{\mu\dot\al\al}\si^\nu_{\al\dot\bet}-
\ov\si^{\nu\dot\al\al}\si^\mu_{\al\dot\bet})$.

The results \susytransi--\susytransiii\ can be rewritten without referring to a
particular ghost picture, in the following form:
\eqn\Susytrans{\eqalign{
[\ Q^I(\eta_I) ,\ V_{\phi^{a,i\pm}}(z,k)\ ]&=
V_{\bar\lambda^{a,J}}(z,\bar v,k)\ \ \ ,\ \ \ \hskip 1.1cm\bar v_{\dot\beta}=
k_\mu\ \eta^\al_I\sigma_{\al\dot\bet}^\mu\ ,\cr
[\ \ov Q^I(\bar\eta_I) ,\ V_{\phi^{a,i\pm}}(z,k)\ ]&=
V_{\lambda^{a,J}}(z,v,k)\ \ \ ,\ \ \ \hskip 1.1cm
v^\bet=k_\mu\ \bar\eta_{I\dot\al}\bar\sigma^{\mu\dot\al\bet}\ ,\cr
[\ Q^I(\eta_I),V_{\lambda^{a,J}}(z,u,k)\ ]&=
(\eta_I u)\ V_{\phi^{a,j\pm}}(z,k) \ ,\cr
[\ \ov Q^I(\bar\eta_I),V_{\bar\lambda^{a,J}}(z,\bar u,k)\ ]&=
(\bar\eta_I \bar u)\ V_{\phi^{a,j\mp}}(z,k) \ ,\cr
[\ Q^I(\eta_I),V_{\bar\lambda^{a,J}}(z,\bar u,k)\ ]&={1\over\sqrt{2}}\
\delta^{IJ}\ V_{A^a}(z,\xi,k)\ \ \ ,\ \ \ \xi^\mu=
\eta_I^\al\ \sigma^\mu_{\al\dot\beta} \bar u^{\dot\beta}\ ,\cr
[\ \ov Q^I(\bar\eta_I),V_{\lambda^{a,J}}(z,u,k)\ ]&={1\over\sqrt{2}}\
\delta^{IJ}\ V_{A^a}(z,\xi,k)\ \ \ ,\ \ \ \,\xi^\mu=
\bar\eta_{I\dot\al}\ \bar\sigma^{\mu\dot\al\beta} u_\beta\ ,\cr
[\ Q^I(\eta_I) ,V_{A^a}(z,\xi,k)\ ]&={1\over\sqrt{2}}\ V_{\lambda^{a,I}}(z,v,k)\ \ \ ,\ \ \
\hskip 5mmv^\bet=\xi_\mu\ k_\nu\
\eta^\al_I\ (\sigma^{\mu\nu})_\al^{\ \bet}\ ,\cr
[\ \ov Q^I(\bar\eta_I) ,V_{A^a}(z,\xi,k)\ ]&=
{1\over\sqrt{2}}\ V_{\bar\lambda^{a,I}}(z,\bar v,k)\ \ \ ,\ \ \ \hskip 5mm\bar v_{\dot\bet}=
\xi_\mu\ k_\nu\ \bar\eta_{I\dot\al}\
(\bar\sigma^{\mu\nu})^{\dot\al}_{\ \dot\bet}\ .}}

\subsec{Contour deformations and SUSY Ward identities}

In this section we consider the action of the SUSY generator \susy\
on a correlation function of $N$ open string states with vertex
operators $V_i(z_i)$ inserted at the boundary of the disk.
We choose a contour $C_\infty$
that surrounds all vertex positions $z_i$ and consider the integral (see Fig. 2):
\eqn\NPOINT{
\Wc:=\oint_{C_\infty} \fc{dw}{2\pi i}\ \eta_I^\al\ \vev{
V^I_\al(w)\ V_1(z_1)\ V_2(z_2)\ldots V_N(z_N)}\ .}
The SUSY current $V^I_\al(w)$ has conformal dimension one. Therefore at infinity
the correlator must behave like $\sim w^{-2}$.
Since the contour $C_\infty$ may be deformed to infinity, we conclude $\Wc=0$.
\ifig\figii{Closed contour $C_\infty$ and $N$ contours $C_l$ around the points
$z_l$.}
{\epsfxsize=0.4\hsize\epsfbox{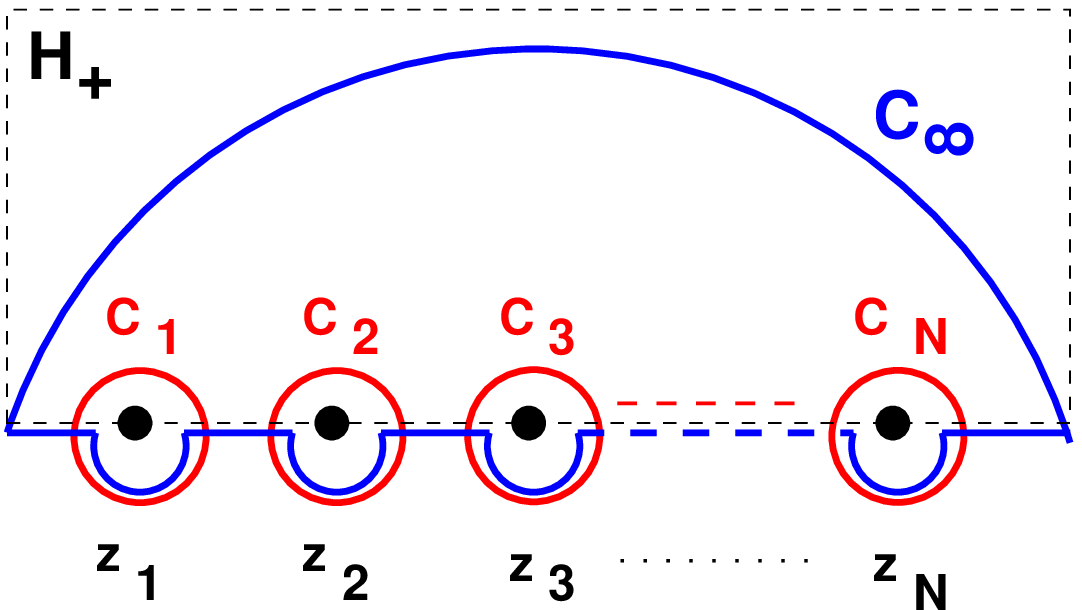}}
\noindent On the other hand, analyticity allows to deform the contour
to a sum over contours $C_l$ encircling each of the points $z_l$ and we may
rewrite \NPOINT\ as:
\eqn\NNPOINT{
\Wc=\sum_{l=1}^N\ \vev{
V_1(z_1)\ldots V_{l-1}(z_{l-1})\ \lf[\oint_{C_l} \fc{dw}{2\pi i}\ \eta_I^\al\
V^I_\al(w)\ V_l(z_l)\ri]\ V_{l+1}(z_{l+1})\ldots V_N(z_N)}\ .}
In the above expression, each SUSY variation \CONTOUR\  gives rise to a
SUSY-transformed vertex operator (depending on $\eta_I^\al$)
discussed in the previous subsection.
Hence the above equation gives rise to
non--trivial relations between different correlators:
\eqn\WARD{
\sum_{l=1}^N\ \vev{
V_1(z_1)\ldots V_{l-1}(z_{l-1})\ [\eta_I^{\al} Q^I_\al,V_l(z_l)]\ V_{l+1}(z_{l+1})
\ldots V_N(z_N)}=0\ .}
Since Eq.\WARD\ holds for arbitrary spinors $\eta$, the coefficients of $\eta_1$ and $\eta_2$ on its l.h.s., as well as the coefficients of $\bar\eta_{\dot 1}$ and $\bar\eta_{\dot 2}$ in the relations involving $\ov Q(\bar\eta)$, must vanish. In this way, SUSY Ward identities can be derived, relating various scattering amplitudes \refs{\ptsusy,\Grisaru,\GrisaruB}. One of such Ward identities will be applied below to the computations of  $N$-gluon amplitudes.
We also use this equation in section 4.2
to relate various four--point string amplitudes.

\subsec{Helicity basis and comparison with $\alpha'\to 0$ limit}
It is well-known from field-theoretical computations that
scattering amplitudes simplify in the so-called helicity basis, that is by considering processes with definite left- and right-handed  polarizations states (helicity -- or +, respectively) of external particles. Furthermore, such a basis is very natural in the context of supersymmetry.
In order to construct the string vertex operators describing helicity eigenstates, it is convenient to express the corresponding wave functions in terms of
two-component, chiral spinors.
We introduce the following shorthand notation for the wave functions of chiral fermions:
\eqn\spin{u^{\al}(k)=k^{\al}\quad,\quad \bar u_{\dot\al}(k)=\bar k_{\dot\al}}
Then the on-shell momentum vectors $k$, with $k^2=0$, factorize as
\eqn\kays{k_{\mu}\sigma^{\mu}_{\al\dot\al}\equiv k_{\al\dot\al}=
k_{\al}\bar k_{\dot\al}\quad,\quad  k^{\mu}\bar\sigma_{\mu}^{\dot\al\al}
\equiv k^{\dot\al\al}= \bar k^{\dot\al} k^{\al}.}
Note that the inverse relation is 
\eqn\ink{k_{\mu}=- {1\over 2}k_{\al\dot\al}\bar\sigma^{\dot\al\al}_{\mu}=- {1\over 2}k^{\al\dot\al}\sigma_{\mu\dot\al\al}\ .}
The bracket products of two spinors associated to momenta $p$ and $q$ are defined as
\eqn\prods{\langle p\, q\rangle=p^{\al}q_{\al}\quad ,\quad[p\, q]=
\bar p_{\dot\al}\bar q^{\dot\al} ~.}
We also use the same symbols for products involving arbitrary spinors:
$\eta^{\al}p_{\al}\equiv \langle\eta p\rangle$ {\it etc}.
In this notation, the polarization vectors of left- and right-handed bosons are \doubref\ManganoBY\DixonWI:
\eqn\polar{\xi^-_{\alpha\dot\alpha}(k,r)=\sqrt{2}\;{k_{\al}\bar
r_{\dot\al}\over[ k\,r]} \quad,\quad \xi^+_{\alpha\dot\alpha}(k,r)=\sqrt{2}\;
{r_{\al}\bar k_{\dot\al}\over\langle r\,k\rangle}~,
}
where $r$ is an arbitrary reference vector.

We introduce the following notation for the vertices describing -- and + helicity states of the ${\cal N}=4$ gauge vector multiplet:\foot{Here, we
skip the
gauge indices.}
\eqn\vern{\eqalign{
&\phi^{i-}(z,k)=V_{\phi^{i-}}(z,k) \qquad \lambda^{J-}(z,k)=
V_{\lambda^J}(z,u^{\al},k)\qquad
g^-(z,k)=V_A(z,\xi^{-\mu},k)\cr
&\phi^{i+}(z,k)=V_{\phi^{i+}}(z,k) \qquad \lambda^{J+}(z,k)=V_{\bar\lambda^J}(z, \bar
u_{\dot\al},k)\qquad g^+(z,k)=V_A(z,\xi^{+\mu},k)\ .}}
Note that for scalars, the above notation is just a matter of convention. Our goal is to simplify the  SUSY transformations \Susytrans\ of these vertices.

First, we consider the transformations of scalars \susytransi. They can
be simplified by using
\eqn\stran{k_{\mu}(\eta^{\al}\sigma_{\al\dot\be}^{\mu}S^{\dot\be})=
\langle\eta k\rangle\, \bar u_{\dot\al}S^{\dot\al}}
and
\eqn\strani{k^{\mu}(\bar\eta_{\dot\al}\,\bar\sigma^{\dot\al\be}_{\mu}S_{\be})=[\eta k] \, u^{\al}S_{\al}}

Next, we consider the gaugino transformations \susytransii. We will prove that
\eqn\gtran{\bar\eta^{\dot\al}u^{\al}(k)\,\sigma^{\mu}_{\al\dot\al}\psi_{\mu}=\sqrt{2}\,[\eta k]\;\xi^{-\mu}\psi_{\mu}}
with the provision that, as in our case, $\psi_{\mu}$ belongs to the gluon vertex operator \fieldsiii\ inserted in an on-shell scattering amplitude. Then the contractions involving $\psi_{\mu}$ amount to replacements $\sigma^{\mu}_{\al\dot\al}\psi_{\mu}\to p_{\al}\bar q_{\dot\al}$, where $p$ and $q$ are some light-like vectors. Then the l.h.s.\ of Eq.\gtran\ becomes
\eqn\gtrani{\bar\eta^{\dot\al}k^{\al}\sigma^{\mu}_{\al\dot\al}\psi_{\mu}=[\eta q]\langle p\, k\rangle.}
On the other hand, after substituting the polarization vector \polar\ into the r.h.s.\ of Eq.\gtran, we obtain
\eqn\gtranii{\sqrt{2}\,[\eta k]\;\xi^{-\mu}\psi_{\mu}=[
k\eta ]{\langle p\, k\rangle [r\, q]\over [k\, r]}=[\eta q]\langle p\, k\rangle  +{[r\,\eta ]\over [k\, r]}\langle p\, k\rangle [k\, q]~,       }
where we used Schouten's identity. After reinstating $\psi_{\mu}$ in the second term on the r.h.s.\ of Eq.\gtranii, $\langle p\, k\rangle [k\, q]\to k^{\mu}\psi_{\mu}$, one finds that it is equivalent to a contribution of the longitudinal part of a polarization vector, therefore it does not contribute to the amplitude. Then Eq.\gtran\ follows from Eqs.\gtrani\ and \gtranii. Similarly, one finds
\eqn\gtranv{\eta^{\al}\bar u^{\dot\al}(k)\,\sigma^{\mu}_{\al\dot\al}\psi_{\mu}
=-\sqrt{2}\langle\eta k\rangle\;\xi^{+\mu}\psi_{\mu}}

Finally, we consider the gluon transformations \susytransiii. The term
\eqn\gltran{\xi_{\mu}k_{\nu}(\eta^{\al}\sigma^{\mu\nu}_{\al\be}S^{\beta})
={1\over
2}(\eta^{\al}k_{\al}\bar k_{\dot\al}\xi^{\dot\al\beta}S_{\be}-\eta^{\al}
\xi_{\al\dot\al}\bar k^{\dot\al} k^{\beta}S_{\be})}
vanishes for the right-handed polarization vector $\xi^+$, {\it c.f}.\ Eq.\polar,
while for the left-handed polarization
\eqn\gltrani{\xi_{\mu}^-k_{\nu}(\eta^{\al}\sigma^{\mu\nu}_{\al\be}S^{\beta})
={\sqrt{2}}\langle \eta k\rangle\, u^{\alpha}(k)S_{\al}}
Similarly, one finds
\eqn\gltranv{\xi_{\mu}^+k_{\nu}(\bar\eta^{\dot\al}
\bar\sigma^{\mu\nu}_{\dot\al\dot\be}
S^{\dot\beta}) =-{\sqrt{2}}[\eta k]\, \bar u_{\dot\alpha}(k)S^{\dot\al}}
Note that all transformations involve one of the factors
\eqn\gams{\Gamma^-(\eta,k)=\langle\eta k\rangle\quad,\quad\Gamma^+(\eta,k)=[\eta k]~.}
After collecting all above formulae, Eq.\Susytrans\ can be rewritten as
\eqn\Susytransi{\eqalign{
[\ Q^I(\eta_I) , {\phi^{i\pm}}(z,k)\ ]&=\Gamma^-(\eta_I, k)\,
{\lambda^{J+}}(z,\bar u,k)\cr
[\ \ov Q^I(\bar\eta_I) , {\phi^{i\pm}}(z,k)\ ]&=\Gamma^+(\eta_I, k)\,
{\lambda^{J-}}(z,u,k)\cr
[\ Q^I(\eta_I),{\lambda^{J-}}(z,k)\ ]&=\Gamma^-(\eta_I, k)\,
{\phi^{j\pm}}(z,k) \ \cr
[\ \ov Q^I(\bar\eta_I),{\lambda^{J+}}(z,k)\ ]&=\Gamma^+(\eta_I, k)\,
{\phi^{j\mp}}(z,k) \ \cr
[\ Q^I(\eta_I),{\lambda^{J+}}(z,k)\ ]&=-\Gamma^-(\eta_I, k)\,
\delta^{IJ} g^+(z,\xi,k)\cr
[\ \ov Q^I(\bar\eta_I),{\lambda^{J-}}(z,k)\ ]&=\Gamma^+(\eta_I, k)\,
\delta^{IJ}g^-(z,\xi,k)\cr
[\ Q^I(\eta_I) ,g^-(z,k)\ ]&=\Gamma^-(\eta_I, k)\,{\lambda^{I-}}(z,k)\cr
[\ \ov Q^I(\bar\eta_I) ,g^+(z,k)\ ]&=-\Gamma^+(\eta_I, k)\,
{\lambda^{I+}}(z,k)~,}}
where the l.h.s.\ and r.h.s.\ indices are paired according to Eq.\Dixon\ and Table 1, keeping in mind that $\Sigma$ and $\ov\Sigma$ are associated to -- and + gaugino helicities, respectively, while $\Psi=\Psi^-$ and $\ov\Psi=\Psi^+$ to -- and + scalars.

As far as practical applications of SUSY relations to the computations of
gluon amplitudes are concerned, there is no need to employ the full ${\cal
N}=4$ algebra. It is often sufficient to consider the gauge multiplet of
${\cal N}=2$ SUSY generated by
$Q^K=(Q^1,i\,Q^2)$ \ptsusy. Such ${\cal N}=2$ SUSY transformations, parameterized by a Dirac spinor $(\eta_K,\bar\eta_K)$, are generated by:
\eqn\strr{Q^K(\eta_K,\bar\eta_K)=\eta^{\al}_KQ^K_{\al}+\bar\eta_{\dot\al K}\ov Q^{\dot\al
K}.}
The ${\cal N}=2$ gauge multiplet consists of the gluon $A$, two gauginos
$\lambda^L=(\lambda^1,i\lambda^2)$  and the scalar $\phi^1\equiv\phi$.
The SUSY transformations of the respective vertex operators can be extracted from Eq.\Susytransi. They are:
\eqn\extend{\eqalign{ &[Q^K(\eta_K,\bar\eta_K),\phi^{\pm}(z,k)]=\pm\,
i\,\varepsilon^{KL} \, \Gamma^{\mp}(\eta_K,k)\,\lambda^{L\pm}(z,k)\ ,\qquad\qquad\cr
&[Q^K(\eta_K,\bar\eta_K),\lambda^{L\pm}(z,k)]=\mp\,\delta^{KL}\,\Gamma^{\mp}
(\eta_K,k)\,g^{\pm}(z,k)\mp\,i\,\varepsilon^{KL}\, \Gamma^{\pm}(\eta_K,k)
\,\phi^{\pm}(z,k)\ ,\cr
&[Q^K(\eta_K,\bar\eta_K),g^{\pm}(z,k)]=\mp\,
\Gamma^{\pm}(\eta_K,k)\lambda^{K\pm}(z,k)\ .\qquad\qquad}}

The field-theoretical relations that should be compared with our SUSY transformations of string vertices are the transformations of the creation and annihilation operators, summarized in Ref.\ptsusy.
Indeed, Eq.\extend\ agree with Eq.(4) of \ptsusy, to all orders in $\al'$.
Since the contour manipulations described at the beginning of this Section are equivalent to applying SUSY Ward identities, all  field-theoretical ($\alpha'=0$) SUSY relations between various amplitudes remain valid in full-fledged  superstring theory.

The main focus of this work are the $N$-gluon MHV disk amplitudes.\comment{\foot{More precisely, the {\it partial} amplitudes associated to one particular gauge group (Chan-Paton) factor ${\rm Tr}  (\, T^{a_1}\cdots T^{a_N})$.}}
%\eqn\mhvg{A_{\rm disk}[\,g^-(k_1),g^-(k_2),g^+(k_3),\dots, g^+(k_N)\,]\ .}
We will use the well-known relation \refs{\ptsusy,\ManganoBY,\DixonWI}
\eqn\mhvg{\eqalign{A(1^-,2^-,3^+,&4^+,\dots,  N^+)=\cr &=
{\langle 1\, 2\rangle^2 \over\langle 3\, 4\rangle^2}\
A[\,\phi^-(k_1),\phi^-(k_2),\phi^+(k_3),\phi^+(k_4),g^+(k_5),\dots, g^+(k_N)\,]~,}}
a consequence of Eqs.\extend,
now guaranteed to hold to all orders in $\alpha'$. It allows replacing
four gluons by four scalars, which is expected to yield a considerable
simplification because the scalar vertices are much easier to handle than
the gluonic ones. We would like to stress that at the disk level,
the $N$-gluon MHV amplitude on the l.h.s.\ of Eq.\mhvg\ is  universal
to all compactifications while the amplitude on the r.h.s.\ involves
the scalar member of ${\cal N}=2$ gauge multiplet, and can be evaluated in an
arbitrary ${\cal N}\ge 2$ compactification. For our purposes, it is most convenient
to use the toroidal, ${\cal N}=4$ compactifications, with the gauge scalar
associated to one of the three complex planes.

We will be first considering the so-called {\it partial} amplitudes, associated to 
one particular gauge group (Chan-Paton) factor ${\rm Tr}  (\, T^{a_1}\cdots T^{a_N})$, 
as in Eq.\mhvform, and later explain how to obtain full amplitudes.

\newsec{Disk scattering of scalars and gluons}

In this section, we compute the string amplitudes for the scattering
of four scalars and $N-4$ gluons at the disk level. All vertex operators are inserted 
at the boundary of the disk. In the notation of Refs.\doubref\STi\STii,
the partial amplitude associated to  ${\rm Tr}  (\, T^{a_1}\cdots T^{a_N})$ Chan-Paton 
factor takes the form:
\eqn\Amp{\eqalign{
&A(\phi^{a_1,-},\phi^{a_2,-},\phi^{a_3,+},\phi^{a_4,+},A^{a_5},\ldots,A^{a_N})=
V_{CKG}^{-1}\int\limits_{z_1<\ldots<z_N}
\lf(\prod_{k=1}^Ndz_k\ri)\cr
&\times \vev{V_{\phi^{a_1,-}}(z_1,k_1)\ V_{\phi^{a_2,-}}(z_2,k_2)\
V_{\phi^{a_3,+}}(z_3,k_3)\ V_{\phi^{a_4,+}}(z_4,k_4)\
\prod_{l=1}^{N-4} V_{A^{a_l}}(z_l,\xi_l,k_l)}\ .}}
The vertex operators for the scalars, in $(-1)-$ and zero--ghost pictures, 
are (\cf also \fieldsi)
\eqn\Vertexop{\eqalign{
V_{\phi^{a,-}}^{(-1)}(z,k)&=g_\phi\ T^a\ e^{-\phi}\ \Psi\ e^{ik_\rho
X^\rho(z)}\ ,\cr
V_{\phi^{a,-}}^{(0)}(z,k)&=\fc{g_\phi}{(2\ap)^{1/2}}\ T^a\
[\ i\p Z+2\ap\ (k\psi)\ \Psi\ ]\ e^{ik_\rho X^\rho(z)}\ ,\cr
V_{\phi^{a,+}}^{(-1)}(z,k)&=g_\phi\ T^a\ e^{-\phi}\ \ov\Psi\ e^{ik_\rho X^\rho(z)}\ ,\cr
V_{\phi^{a,+}}^{(0)}(z,k)&=\fc{g_\phi}{(2\ap)^{1/2}}\ T^a\
[\ i\p \ov Z+2\ap\ (k\psi)\ \ov\Psi\ ]\ e^{ik_\rho X^\rho(z)}\ ,}}
respectively.
The vertex operator for the gauge boson
has been already written before in the $(-1)$--ghost picture in Eq.\fieldsiii,  
and in the zero--ghost picture it takes the form:
\eqn\gaugevertexzero{
V_{A^a}^{(0)}(z,\xi,k)=\fc{g_A}{(2\ap)^{1/2}}\
T^a\xi_\mu\ [\ i\p X^\mu+2\ap\ (k\psi)\ \psi^\mu\ ]\ e^{ik_\rho X^\rho(z)}\ .}
In order to
cancel the background ghost charge on the disk, two vertices in the correlator \Amp\ 
will be inserted in the $(-1)$--ghost picture, with
the remaining ones  in the zero--ghost picture.
Furthermore, in Eq.\Amp, the factor $V_{CKG}$
accounts for the volume of the
conformal Killing group of the disk after choosing the conformal gauge.
It will be canceled by fixing three vertex positions and introducing the
respective $c$--ghost correlator.
The correlator of vertex operators in \Amp\ is evaluated by performing
all possible Wick contractions.
It decomposes into products of the two--point functions on the
boundary of the disk:
\eqn\green{\eqalign{
&\vev{\p X^\mu(z_1) X^\nu(z_2)}=-\fc{2\ap\delta^{\mu\nu}}{z_{12}}\ ,\
\vev{\p X^\mu(z_1) \p X^\nu(z_2)}=-\fc{2\ap\delta^{\mu\nu}}{z_{12}^2}\ ,\ \cr
&\vev{e^{ik_\mu X^\mu(z_1)} e^{ik_\nu X^\nu(z_2)}}=|z_{12}|^{2\ap k_1k_2}\ ,\
\vev{e^{-\phi(z_1)}e^{-\phi(z_2)}}=\fc{1}{z_{12}}\ ,\
\vev{\psi^\mu(z_1)\psi^\nu(z_2)}=\fc{\delta^{\mu\nu}}{z_{12}}\ ,\cr
&\vev{\Psi(z_1)\ov\Psi(z_2)}=\fc{1}{z_{12}}\ ,\
\vev{\Psi(z_1)\Psi(z_2)}=0\ ,\cr
&\vev{\p Z(z_1)\p\ov Z(z_2)}=-\fc{2\ap}{z_{12}^{2}}\ ,\
\vev{\p Z(z_1)\p Z(z_2)}=0\ .}}
Because of the $PSL(2,\IR)$ invariance on the disk, we can fix three positions
of the vertex operators. A convenient choice respecting the
integration region $z_1<\ldots<z_N$ is
\eqn\choice{
z_1=-z_\infty=-\infty\ \ \ ,\ \ \ z_2=0\ \ \ ,\ \ \ z_3=1\ ,}
which implies the ghost factor $\vev{c(z_1)c(z_2)c(z_3)}=-z_\infty^2$.
The remaining $N-3$ vertex positions $z_4,\ldots z_N$ take arbitrary values inside
the integration domain $1<z_4<\ldots<z_N<~\infty$.

In order to correctly normalize the amplitudes, some additional factors
%of the bosonic correlators in \green\ and the ghost correlator
have to be
taken into account.  They stem from determinants
and Jacobians of certain path integrals. On the disk, the net result of those
contributions is an additional factor of 
\eqn\diskfactor{C_{D_2}={1\over 2\, g_{Y\! M}^2\,\ap^2}\ ,}
which must be included in all disk correlators \JOE.

\subsec{Five--point amplitudes}

{}For $N=5$, Eq.\Amp\ yields the following correlator:
\eqn\AmpV{
\vev{V_{\phi^{a_1,-}}^{(0)}(z_1,k_1)\ V_{\phi^{a_2,-}}^{(0)}(z_2,k_2)\
V_{\phi^{a_3,+}}^{(-1)}(z_3,k_3)\ V_{\phi^{a_4,+}}^{(0)}(z_4,k_4)\
V^{(-1)}_{A^{a_5}}(z_5,\xi_5,k_5)}\ .}
Proceeding as outlined above, we obtain
\eqn\arrive{\eqalign{
A(\phi^{a_1,-},\phi^{a_2,-},\phi^{a_3,+},\phi^{a_4,+},A^{a_5})&=2(2\ap)\ g_{Y\! M}^3
\ \Tr(T^{a_1}T^{a_2}T^{a_3}T^{a_4}T^{a_5})\
V_{CKG}^{-1}\cr
&\hskip-3.5cm\times\int\limits_{z_1<\ldots<z_5}\
\lf(\prod_{k=1}^5 dz_k\ri)\lf(\prod_{i<j}|z_{ij}|^{s_{ij}}\ri)\
\fc{1}{z_{35}}\ \lf\{\fc{(\xi_5k_4)}{z_{45}}\
\fc{s_{12}\ z_{34}}{z_{24}z_{13}z_{14}z_{23}}\ri.\cr
&\hskip-2.9cm+\lf.
\fc{(\xi_5 k_1)}{z_{15}z_{24}}\
\lf[\fc{(1-s_{24})}{z_{24}z_{13}}+
\fc{s_{24}}{z_{14}z_{23}}\ri]+\fc{(\xi_5 k_2)}{z_{14}z_{25}}\
\lf[\fc{(1-s_{14})}{z_{14}z_{23}}+
\fc{s_{14}}{z_{13}z_{24}}\ri]\ \ri\}\ ,}}
with $s_{ij}\equiv 2\ap k_ik_j$ \doubref\STi\STii.
We set $z_1,z_2,z_3$ as in \choice\ (taking into account the ghost factor)
and use the  parameterization
\eqn\usepara{
z_4=x^{-1}\ \ \ ,\ \ \ z_5=(x\ y)^{-1}\ ,\qquad (0<x~,~y<1)~, }
with the corresponding Jacobian
$\det(\fc{\p(z_4,z_5)}{\p(x,y)})=x^{-3}y^{-2}$. Then Eq.\arrive\ becomes
\eqn\becomes{\eqalign{
A(\phi^{a_1,-},\phi^{a_2,-},\phi^{a_3,+},\phi^{a_4,+},A^{a_5})&=
2(2\ap)\ g_{Y\! M}^3\ \Tr(T^{a_1}T^{a_2}T^{a_3}T^{a_4}T^{a_5})\cr
&\times\lf[\ (\xi_5k_1)\ K_1
+(\xi_5k_2)\ K_2+(\xi_5k_4)\  K_3\ \ri]\ ,}}
with the three functions
\eqn\functions{\eqalign{
K_1&=\int\limits_0^1 dx\int\limits_0^1 dy\
\lf(1-s_{24}+\fc{s_{24}}{x}\ri)\ \fc{\Ic(x,y)}{y(1-xy)}\ ,\cr
K_2&=\int\limits_0^1 dx\int\limits_0^1 dy\
\lf(\fc{1-s_{14}}{x}+s_{14}\ri)\ \fc{\Ic(x,y)}{(1-xy)}\ ,\cr
K_3&=s_{12}\ \int\limits_0^1 dx\int\limits_0^1 dy\
\fc{(1-x)}{x(1-y)}\ \fc{\Ic(x,y)}{(1-xy)}\ ,}}
where
\eqn\integrand{
\Ic(x,y)=x^{s_2}\ y^{s_5}\ (1-x)^{s_3}\ (1-y)^{s_4}\ (1-xy)^{s_1-s_3-s_4}\ ,}
and $s_i\equiv\ap(k_i+k_{i+1})^2$, subject to the cyclic identification
$i+5\equiv i$.
The functions \functions\
integrate to Gaussian hypergeometric functions $\FF{3}{2}$ \STii.
Actually, $K_3$ may be expressed in terms of  $K_1$ and $K_2$:
\eqn\maybeexpressed{
K_3=-\fc{s_5}{s_4}\ K_1+\fc{s_1-s_3+s_5}{s_4}\ K_2\ .}
Thus as expected, $\ap$-dependence of the amplitude describing the scattering of four-scalars and one gluon is determined by two independent functions, exactly as many
as in the five-gluon case \threeref\Brazil\Dan\STii. 

The low--energy behavior of the amplitude \becomes\ is determined, up to the
order $\Oc(\ap^2)$, by the following expansions:
\eqn\expansions{\eqalign{
K_1&=\fc{1}{s_2}-\fc{s_3}{s_2s_5}-\zeta(2)\
\lf(s_1+s_3-\fc{s_3s_4}{s_2}+\fc{s_4s_5}{s_2}-\fc{s_3^2}{s_5}\ri)+\ldots\ ,\cr
K_2&=\fc{1}{s_2}-\zeta(2)\ \lf(s_1+s_3+\fc{s_4s_5}{s_2}\ri)+\ldots\ ,\cr
K_3&=\fc{s_1}{s_2s_4}-\zeta(2)\ \lf(\fc{s_1s_5}{s_2}+\fc{s_1^2}{s_4}\ri)+\ldots\ .}}

The functions $\{K_1,K_2\}$ can be expressed in terms of another two-element
basis, previously used in \doubref\STi\STii.
There, we defined two functions:
\eqn\funct{
f_1=\int\limits_0^1 dx\int\limits_0^1 dy\
x^{-1}\ y^{-1}\ \Ic(x,y)\ \ \ \ ,\ \ \
f_2=\int\limits_0^1 dx\int\limits_0^1 dy\ (1-xy)^{-1}\ \Ic(x,y)\ .}
Expressed in terms of  $\{f_1,f_2\}$, the functions $K_1$ and $K_2$ read:
\eqn\read{
K_1=(s_5-s_3)\ f_1-s_1\ f_2\ \ \ ,\ \ \ K_2=s_5\ f_1-s_1\ f_2\ .}

The result \becomes\ can be further simplified by choosing $k_4$ as the
reference  vector $r$ for the polarization vector $\xi_5$, so that $\xi_5k_4=0$, see
Eq.\polar. 
Then, for the positive polarization of the gluon,
\eqn\epsk{\xi^+_5k_1=-{1\over\sqrt{2}}{\langle 4\, 1\rangle[1\, 5]\over\langle
4\, 5\rangle}\quad,\quad
\xi^+_5k_2=-{1\over\sqrt{2}}{\langle 4\, 2\rangle[2\, 5]\over\langle 4\,
5\rangle}\ .}
As a result of combining Eq.\becomes\ with the SUSY relation \mhvg, after 
factorizing out $(\sqrt{2}\,g_{Y\! M})^3
\, \Tr(T^{a_1}\cdots T^{a_5})$, we obtain the five-gluon MHV amplitude:
\eqn\fivegl{A(1^-,2^-,3^+,4^+,5^+)=\ap {\langle 1\, 2\rangle^2\over\langle 3\,
4\rangle^2\langle 4\, 5\rangle}(\, \langle 4\, 1\rangle[1\, 5]\ K_1+
\langle 4\, 2\rangle[2\, 5]\ K_2\, )\ .}
The above result should be compared with the factorized form derived in \STii
\eqn\mfive{
A(1^-,2^-,3^+,4^+,5^+)=
[\ V^{(5)}(s_j)-2i\
P^{(5)}(s_j)\ \epsilon(1,2,3,4)\ ]\ {\cal M}_{Y\!M}^{(5)}\ ,}
where:
\eqn\vfive{
V^{(5)}(s_i)= s_2s_5\ f_1 + {1\over 2} (s_2s_3+s_4s_5-s_1s_2-s_3s_4-s_1s_5)\ f_2
\quad,\quad P^{(5)}(s_i)= f_2 \ .}
Indeed, it is a matter of simple spinor manipulations to show that Eqs.\fivegl\ and
\mfive\ agree upon using the relations \read\ between the two sets of basis functions,
$\{K_1, K_2\}$ and $\{f_1, f_2\}$. Unlike the factorized amplitude \mfive, the new form \fivegl\
is expressed directly in terms of ``primitive'' integrals \functions, without resorting to
the definitions of  ``formfactor'' functions $V$ and $P$, see Eq.\vfive. Although for five gluons it is not a dramatic simplification, we will see that for six gluons and more, the non-factorized form is more suitable.

\subsec{Six--point amplitudes}

Now we turn to  $N=6$. For that case in \Amp, we compute the
following correlator:
\eqn\AmpVI{
\vev{V_{\phi^{a_1,-}}^{(0)}(z_1,k_1)V_{\phi^{a_2,-}}^{(0)}(z_2,k_2)
V_{\phi^{a_3,+}}^{(0)}(z_3,k_3)V_{\phi^{a_4,+}}^{(0)}(z_4,k_4)
V^{(-1)}_{A^{a_5}}(z_5,\xi_5,k_5)V^{(-1)}_{A^{a_6}}(z_6,\xi_6,k_6)}\ .}
With the choice \choice\ and the  parameterization
\eqn\Usepara{
z_4=x^{-1}\ \ \ ,\ \ \ z_5=(x\ y)^{-1}\ \ \ ,\ \ \ z_6=(x\ y\ z)^{-1}\ ,}
the Jacobian
$\det(\fc{\p(z_4,z_5,z_6)}{\p(x,y,z)})=x^{-4}y^{-3}z^{-2}.$
The amplitude \AmpVI\ becomes
\eqn\Becomes{\eqalign{
A(&\phi^{a_1,-},\phi^{a_2,-},\phi^{a_3,+},\phi^{a_4,+},A^{a_5},A^{a_6})=
2(2\ap)^2\ g_{Y\! M}^4\ \Tr(T^{a_1}T^{a_2}T^{a_3}T^{a_4}T^{a_5}T^{a_6})\cr
&\times\lf[(\xi_5k_3)(\xi_6k_2)\ K_1
+(\xi_5k_2)(\xi_6k_3)\ K_2+
(\xi_5k_1)(\xi_6k_2)\ K_3+
(\xi_5k_1)(\xi_6k_3)\ K_4\ri.\cr
&~+(\xi_5k_2)(\xi_6k_1)\ K_5+
(\xi_5k_3)(\xi_6k_1)\ K_6+
(\xi_5k_3)(\xi_6k_4)\ K_7+
(\xi_5k_4)(\xi_6k_3)\ K_8\cr
&~+(\xi_5k_2)(\xi_6k_4)\ K_9+(\xi_5k_4)(\xi_6k_2)\ K_{10}
+(\xi_5k_1)(\xi_6k_4)\ K_{11}+(\xi_5k_4)(\xi_6k_1)\ K_{12}\cr & \lf.
~+\,(\xi_5\xi_6)\ K_{13}\ \ri]\ ,}}
where the integrals
\eqn\Functions{\eqalign{
K_1&=\int\limits_0^1 dx\int\limits_0^1 dy\ \int\limits_0^1 dz\
\lf(\fc{1-s_{14}}{x}+s_{14}\ri)\ \fc{\Ic(x,y,z)}{(1-z)(1-xy)}\ ,\cr
K_2&=-\int\limits_0^1 dx\int\limits_0^1 dy\ \int\limits_0^1 dz\
\lf(\fc{1-s_{14}}{x}+s_{14}\ri)\ \fc{\Ic(x,y,z)}{(1-z)(1-xyz)}\ ,\cr
K_3&=s_{34}\ \int\limits_0^1 dx\int\limits_0^1 dy\ \int\limits_0^1 dz\
\fc{\Ic(x,y,z)}{xy(1-z)}\ ,\cr
K_4&=-\int\limits_0^1 dx\int\limits_0^1 dy\ \int\limits_0^1 dz\
\lf(1-s_{24}+\fc{s_{24}}{x}\ri)\ \fc{\Ic(x,y,z)}{y(1-z)(1-xyz)}\ ,\cr
K_5&=-s_{34}\ \int\limits_0^1 dx\int\limits_0^1 dy\ \int\limits_0^1 dz\
\fc{\Ic(x,y,z)}{xyz(1-z)}\ ,\cr
K_6&=\int\limits_0^1 dx\int\limits_0^1 dy\ \int\limits_0^1 dz\
\lf(1-s_{24}+\fc{s_{24}}{x}\ri)\ \fc{\Ic(x,y,z)}{yz(1-z)(1-xy)}\ ,}}
and the remaining seven integrals are displayed in Eq.(B.1) of Appendix \appB.
All integrands contain the common factor
\eqn\Integrand{\eqalign{
\Ic(x,y,z)&=x^{s_2}\ y^{t_2}\ z^{s_6}\
(1-x)^{s_3}\ (1-y)^{s_4}\ (1-z)^{s_5}\cr
&\times(1-xy)^{t_3-s_3-s_4}\ (1-yz)^{t_1-s_4-s_5}\ (1-xyz)^{s_1+s_4-t_1-t_3}\ ,}}
where $s_i=\ap(k_i+k_{i+1})^2$ and $t_j=\ap(k_j+k_{j+1}+k_{j+2})^2$,
subject to the cyclic identification $i+6\equiv i$.

The integrals \Functions\ and (B.1) integrate to multiple Gaussian
hypergeometric functions, more
precisely they represent triple hypergeometric functions \Dan.
Although the result \Becomes\ uses thirteen functions, only six of them are linearly independent, exactly as many as in the  six-gluon case  \Dan. {}For our purposes, it is convenient to choose $\{K_1,\ldots,K_6\}$ as the basis. The remaining seven functions are then expressed in terms of these functions in Eq.(B.2)

The low--energy behavior of the amplitude \becomes\ is determined, up to the
order $\Oc(\ap^2)$, by the following expansions:
\eqn\Expansions{\eqalign{
K_1&=\fc{1}{s_2s_5}+\zeta(2)\
\lf(1-\fc{s_4+s_5+s_6-t_1-t_2}{s_2}-\fc{s_1+s_3}{s_5}-\fc{t_1t_2}{s_2s_5}\ri)+
\ldots\ ,\cr
K_2&=-\fc{1}{s_2s_5}+\zeta(2)\
\lf(\fc{s_4+s_5+s_6-t_1-t_2}{s_2}+\fc{s_1+s_3}{s_5}+\fc{t_1t_2}{s_2s_5}\ri)+
\ldots\ ,\cr
K_3&=\fc{s_3}{s_2s_5t_2}+\zeta(2)\ \lf(\fc{s_3}{s_2}-\fc{s_3t_1}{s_2s_5}-
\fc{s_3^2}{s_5t_2}-\fc{s_3s_6}{s_2t_2}\ri)+\ldots\ ,\cr
K_4&=-\fc{1}{s_2s_5}+\fc{s_3}{s_2s_5t_2}+\zeta(2)
\lf(\fc{s_3+s_6-t_2}{s_2}+\fc{s_1+s_3}{s_5}-\fc{s_3t_1}{s_2s_5}-\fc{s_3^2}{s_5t_2}
-\fc{s_3s_6}{s_2t_2}+\fc{t_1t_2}{s_2s_5}\ri)+\ldots\ ,\cr
K_5&=-\fc{s_3}{s_2s_5t_2}-\fc{s_3}{s_2s_6t_2}+\zeta(2)\lf(-\fc{s_3}{s_2}+
\fc{s_3s_4}{s_2s_6}+\fc{s_3t_1}{s_2s_5}+\fc{s_3^2}{s_5t_2}+\fc{s_3s_5}{s_2t_2}+
\fc{s_3^2}{s_6t_2}+\fc{s_3s_6}{s_2t_2}\ri)+\ldots\ ,\cr
K_6&=\fc{1}{s_2s_5}+\fc{1}{s_2s_6}-\fc{s_3}{s_2s_5t_2}-\fc{s_3}{s_2s_6t_2}+
\zeta(2)\lf(1-\fc{s_3+s_5+s_6+t_2}{s_2}-\fc{s_1+s_3}{s_5}-\fc{s_3+t_3}{s_6}\ri.\cr
&\hskip2cm
\lf.+\fc{s_3s_4}{s_2s_6}+\fc{s_3t_1}{s_2s_5}+
\fc{s_3^2}{s_5t_2}+\fc{s_3s_5}{s_2t_2}+
\fc{s_3^2}{s_6t_2}+\fc{s_3s_6}{s_2t_2}-\fc{s_4t_2}{s_2s_6}-\fc{t_1t_2}{s_2s_5}\ri)+
\ldots\ .}}

At this point, we specify the result \Becomes\ to the case of two positive-helicity gluons, as in the
amplitude entering the SUSY relation \mhvg.
We choose $k_4$ as the reference vector $r$ for both polarization vectors, so that $\xi_ik_4=0$, see Eq.\polar. Then:
\eqn\epsk{ \xi^+_i\xi_j^+=0    \quad,\quad
\xi^+_ik_j=-{1\over\sqrt{2}}{\langle 4\, j\rangle[j\, i]\over\langle 4\,
i\rangle}\ .}
We also define:
\eqn\defxi{
\tau(a,b)=2\,(\xi_5^+k_a)\ (\xi_6^+k_b)={\langle 4\, a\rangle\langle 4\,
b\rangle[a\, 5][b\, 6]\over\langle 4\, 5\rangle\langle 4\, 6\rangle }\ .}
As a result of combining Eqs.\AmpVI\ and \mhvg, after factorizing out 
$(\sqrt{2}\,g_{Y\! M})^4\, \Tr(T^{a_1}\cdots T^{a_6})$,
we obtain the six-gluon MHV amplitude:
\eqn\sixgl{\eqalign{A(1^-,2^-,3^+,4^+,5^+, 6^+)=\ap^2{\langle 1\, 
2\rangle^2\over\langle 3\, 4\rangle^2}\left[\,
\tau(3,2)\ K_1+\tau(2,3)\ K_2+\tau(1,2)K_3\right. ~~~~\cr\left.
+\,\tau(1,3)\ K_4+\tau(2,1)\ K_5+\tau(3,1)\ K_6\,\right]\ . }}
The above result should be compared with the factorized form derived in \STii:
\eqn\six{A(1^-,2^-,3^+,4^+,5^+,6^+)=
\lf[\ V^{(6)}(s_i,t_i)-2i\ \sum_{k=1}^{k=5}\,\epsilon_k\,P^{(6)}_k(s_i,t_i)\ \ri]\
{\cal M}_{Y\!M}^{(6)}}
where
\eqn\sixeps{\epsilon_1=\epsilon(2,3,4,5)
\quad\epsilon_2=\epsilon(1,3,4,5)
\quad\epsilon_3=\epsilon(1,2,4,5)\quad\epsilon_4=\epsilon(1,2,3,5)
\quad\epsilon_5=\epsilon(1,2,3,4).}
The functions $V^{(6)}$ and $ P^{(6)}_k, k=1,\dots, 5,$ \doubref\STi\STii\ are certain (complicated) linear combinations of:
\eqn\Functs{\eqalign{
&F_1=\INT\fc{\Ic(x,y,z)}{xyz}\qquad \hskip 1.3cm F_2=\INT\fc{\Ic(x,y,z)}{z\ (1-xy)}\cr
&F_3=\INT\fc{\Ic(x,y,z)}{1-xyz}\qquad \hskip 1.3cm
F_4=\INT\fc{y\ \Ic(x,y,z)}{(1-xy)(1-yz)}\cr
&F_5=\INT\fc{\Ic(x,y,z)}{(1-xy)(1-xyz)}\qquad
F_6=\INT\fc{\Ic(x,y,z)}{(1-yz)(1-xyz)}\ ,}}
weighted by coefficients which are polynomial in the kinematic invariants.

In order to compare Eqs.\sixgl\ and \six, one needs relations between $K$- and $F$-functions.
These can be obtained by integrating by parts combined with other manipulations described in
\STii. One finds, for instance,
\eqn\Read{\eqalign{
s_2\ s_5\ K_3&=s_2 s_3 s_6\ F_1+s_3 s_6 (s_4+s_5-t_1)\ F_2
+s_3 (s_4+s_5-t_1) (s_1+s_3-s_5-t_3)\ F_6\cr
&+s_3 \lf[-s_1 s_2-
(s_4+s_5-t_1)\ (s_3-s_5+s_6+t_1-t_3)+s_2 (s_5+t_3)\ri]\ F_3\cr
&+s_3 (s_4+s_5-t_1) (-s_3+s_5-t_1+t_3)\ F_4+s_3 (s_4+s_5-t_1)
(-s_1+s_3-s_5+t_1) F_5\ .}}
The remaining five elements of the $K$-basis are expressed in terms of
$\{F_1,F_2,F_3,F_4,F_5,F_6\}$ in Appendix \appB. After some tedious
algebra one finds that the new expression \sixgl\ does indeed agree with the
results of \doubref\STi\STii. The use of $K$-basis imposed by SUSY relations leads to a
much simpler expression, as a linear combination of ``primitive'' integrals
\Functions\ weighted by single-term twistor-like coefficients.

\subsec{Seven--point amplitudes}

Finally, we consider the case of $N=7$.
We evaluate the following correlator:
\eqn\AmpVII{
\vev{V_{\phi^{a_1,-}}^{(0)}(z_1)V_{\phi^{a_2,-}}^{(0)}(z_2)
V_{\phi^{a_3,+}}^{(0)}(z_3)V_{\phi^{a_4,+}}^{(0)}(z_4)
V^{(-1)}_{A^{a_5}}(z_5)V^{(-1)}_{A^{a_6}}(z_6)
V^{(0)}_{A^{a_7}}(z_7)}\ .}
Here again, we make the choice \choice\ and use the  parameterization
\eqn\Useparaa{
z_4=x^{-1}\ \ \ ,\ \ \ z_5=(x\ y)^{-1}\ \ \ ,\ \ \ z_6=(x\ y\ z)^{-1}\ \ \ ,\
\ \ z_7=(x\ y\ z\ w)^{-1}\ ,}
with the corresponding Jacobian
$\det(\fc{\p(z_4,z_5,z_6,z_7)}{\p(x,y,z,w)})=x^{-5}y^{-4}z^{-3}w^{-2}$.
Even after using momentum conservation to eliminate the scalar
products $(\xi_5k_6),\ (\xi_6k_7)$ and $(\xi_7k_5)$,
\eqn\impose{\eqalign{
\xi_5k_6&=-\xi_5k_1-\xi_5k_2-\xi_5k_3-\xi_5k_4-\xi_5k_7\ ,\cr
\xi_6k_7&=-\xi_6k_1-\xi_6k_2-\xi_6k_3-\xi_6k_4-\xi_6k_5\ ,\cr
\xi_7k_5&=-\xi_7k_1-\xi_7k_2-\xi_7k_3-\xi_7k_4-\xi_7k_6\ ,}}
the amplitude \AmpVII\ still involves many\foot{After applying \impose\
the amplitude \AmpVII\
involves $111$ different kinematical factors.
Each of them is multiplied by a certain integral or hypergeometric function.
However, some of those kinematical factors appear with the same function.
In fact, only $73$ different integrals or functions appear in \AmpVII.
These $73$ functions may be expressed in terms of a basis,
whose dimension is $24$. As in the five or six--point case this basis is
completely specified by the MHV part of the amplitude \AmpVII.}
kinematic factors, each of
them multiplied by a certain Euler integral or multiple hypergeometric
function. However, all integrals can
be expressed in terms of the 24-element basis $\{K_1,\dots, K_{24}\}$
written in Appendix \appC.
As we will see below, similarly to the six-gluon case, this basis
appears naturally  in the MHV part of the amplitude \AmpVII.

At this point, we specify the correlator \AmpVII\ to
the case of three positive--helicity gluons, as in the
amplitude entering the SUSY relation \mhvg.
We choose $k_4$ as the reference vector $r$ for all polarization vectors, so
that $\xi_ik_4=0$, see Eq. \polar. Then:
\eqn\epsk{ \xi^+_i\xi_j^+=0    \quad,\quad
\xi^+_ik_j=-{1\over \sqrt{2}}{\langle 4\, j\rangle[j\, i]\over\langle 4\, i\rangle}\ .}
We also define:
\eqn\defxsi{
\tau(a,b,c)=-2\sqrt{2}(\xi_5k_a)\ (\xi_6k_b)\ (\xi_7k_c)={\langle 4\, a\rangle\langle
4\, b\rangle\langle 4\, c\rangle[a\, 5][b\, 6][c\, 7]\over\langle 4\,
5\rangle\langle 4\, 6\rangle\langle 4\, 7\rangle\ }\ .}
For $\xi_ik_4=0$, the correlator \AmpVII\ gives
rise to the partial amplitude
\eqn\Becomess{\eqalign{
A(\phi^{a_1,-},\phi^{a_2,-},\phi^{a_3,+},
\phi^{a_4,+},A^{a_5},A^{a_6},A^{a_7})&=2(2\ap)^3g_{Y\! M}^5\cr
&\times\Tr(T^{a_1}T^{a_2}T^{a_3}T^{a_4}T^{a_5}T^{a_6}T^{a_7})\
\sum_{I=1}^{24} \Kc_I\ K_I\ ,}}
with $24$ kinematics $\Kc_I$ and integrals $K_I$ to be specified below.
After using the SUSY relation \mhvg\ and factorizing out $(\sqrt{2}\,g_{Y\! M})^5
\, \Tr(T^{a_1}\cdots T^{a_7})$, the correlator \AmpVII\ yields the
seven--gluon MHV amplitude
\eqn\Becomess{A(1^-,2^-,3^+,4^+,5^+,6^+,7^+)=\ap^3{\langle 1\,
2\rangle^2\over\langle 3\, 4\rangle^2}~
\sum_{I=1}^{24} \Kc_I\ K_I\ ,}
where the kinematic factors  $\Kc_I$ are as follows:
\eqn\twentyfour{\eqalign{
\Kc_1&=\tau(1,2,1)+\tau(7,2,1)\ \ ,\ \ \Kc_2=\tau(1,3,1)+\tau(7,3,1)\ \ ,\ \
\Kc_3=\tau(2,2,1)+\tau(2,5,1)\ ,\cr
\Kc_4&=\tau(3,3,1)+\tau(3,5,1)\ \ ,\ \
\Kc_5=\tau(2,1,2)+\tau(7,1,2)\ \ ,\ \
\Kc_6=\tau(2,3,2)+\tau(7,3,2)\ ,\cr
\Kc_7&=\tau(1,1,2)+\tau(1,5,2)\ \ ,\ \
\Kc_8=\tau(3,3,2)+\tau(3,5,2)\ \ ,\ \
\Kc_9=\tau(3,1,3)+\tau(7,1,3)\ ,\cr
\Kc_{10}&=\tau(3,2,3)+\tau(7,2,3)\ \ ,\ \
\Kc_{11}=\tau(1,1,3)+\tau(1,5,3)\ \ ,\ \
\Kc_{12}=\tau(2,2,3)+\tau(2,5,3)\ ,\cr
\Kc_{13}&=\tau(2,1,1)+\tau(2,1,6)\ \ ,\ \
\Kc_{14}=\tau(3,1,1)+\tau(3,1,6)\ \ ,\ \
\Kc_{15}=\tau(1,2,2)+\tau(1,2,6)\ ,\cr
\Kc_{16}&=\tau(3,2,2)+\tau(3,2,6)\ \ ,\ \
\Kc_{17}=\tau(1,3,3)+\tau(1,3,6)\ \ ,\ \
\Kc_{18}=\tau(2,3,3)+\tau(2,3,6)\ ,\cr
\Kc_{19}&=\tau(3,2,1)\quad ,\qquad\hskip 11mm \Kc_{20}=\tau(2,3,1)\quad,\qquad\hskip 11mm
~\Kc_{21}=\tau(3,1,2),\ \cr \Kc_{22}&=\tau(1,3,2)\quad ,\qquad\hskip 11mm
\Kc_{23}=\tau(2,1,3)\quad ,\qquad\hskip 11mm ~\Kc_{24}=\tau(1,2,3)\ .}}
The $24$ basis functions $K_i$ are given as Euler integrals, e.g.:
\eqn\egg{
K_1=-s_{34}\ \int\limits_0^1 dx\int\limits_0^1 dy\ \int\limits_0^1 dz\ \int\limits_0^1
dw\ \fc{\Ic(x,y,z,w)}{xyw(1-z)(1-wz)}\ ,}
with
\eqn\Integrandd{\eqalign{
\Ic(x,y,z,w)&=x^{s_2}\ y^{t_2}\ z^{t_6}\ w^{s_7}
(1-x)^{s_3}\ (1-y)^{s_4}\ (1-z)^{s_5}\ (1-w)^{s_6}\cr
&\times(1-xy)^{t_3-s_3-s_4}\ (1-yz)^{t_4-s_4-s_5}\
(1-xyz)^{s_4-t_3-t_4+t_7}\cr
&\times (1-zw)^{t_5-s_5-s_6}\ (1-yzw)^{s_5+t_1-t_4-t_5}\
(1-xyzw)^{s_1-t_1+t_4-t_7}\ ,}}
$s_i=\ap(k_i+k_{i+1})^2$ and $t_j=\ap(k_j+k_{j+1}+k_{j+2})^2$
subject to the cyclic identification $i+7\equiv i$. All $24$ integrals are
displayed in Appendix \appC.
The $18$ functions $K_1\ldots,K_{18}$ seem like straightforward generalizations
of the six-point functions \Functions. Indeed, they
are related to them in certain soft-boson limits:
{}for example, in the $k_7\to 0$ limit, $K_1\to -\fc{1}{s_7}K_3$  of $N=6$.
On the other hand, the  remaining functions $K_{19},\ldots,K_{24}$ seem
to have a different character.
\subsec{From partial to full amplitudes}
So far, we focussed only on one partial amplitude, associated to the
color factor $\Tr(T^{a_1}\cdots T^{a_N})$. In Ref.\STii,  we gave a prescription for constructing any partial amplitude $A_{\sigma}(1^-,2^-,3^+,4^+,\dots, N^+)$, associated
the color factor $\Tr (\, T^{a_{\sigma(1)}}\cdots T^{a_{\sigma(N)}})$ with the indices
$a_i$ permuted by an arbitrary permutation $\sigma$.
One simply factorizes out $\langle 1\,2\rangle^4$ and applies $\sigma$ to all momenta inside the remainder:
\eqn\other{A_{\sigma}(1^-,2^-,3^+,4^+,\dots, N^+)=\langle 1\,2\rangle^4\times\left(\left. {A(1^-,2^-,3^+,4^+,\dots, N^+)\over\langle 1\,2\rangle^4} \right|_{k_i\to k_{\sigma(i)}}\right).}
{} For an amplitude written as in \STii, as the product of the zero-slope MHV amplitude ${\cal M}_{Y\!M}^{(N)}$
\mhvpt\ times a string ``formfactor'', this is a completely trivial operation, however
with the non-factorized form of amplitudes presented in this section, it requires more care.

\newsec{Four--point string amplitudes and supersymmetry relations}

In the previous section we have seen, that string amplitudes may be
written much simpler by making use of the  SUSY Ward identities \WARD.
In this Section we demonstrate this for the four--point string amplitude
involving scalars, gauginos and vectors from the $\Nc=4$ vector multiplet.
After computing the latter we apply the results of Section 2.3 to generate
relations among them.

\subsec{Four--point disk scattering of scalars, gauginos and gluons}

Here we compute in $D=4$  the tree--level four--point string amplitude
involving scalars, gauginos and vectors from the $\Nc=4$ vector multiplet.
The world--sheet of the string
$S$--matrix is described by a disk with all external states $\Phi^a$
created through vertex operators $V_{\Phi^a}$ at the boundary of the disk.
The color--ordered part of the amplitude of interest takes the form:
\eqn\Vier{
A(\Phi^{a_1},\Phi^{a_2},\Phi^{a_3},\Phi^{a_4})=
V_{CKG}^{-1}\int\limits_{z_1<\ldots<z_4}
\lf(\prod_{k=1}^4 dz_k\ri)\ \vev{V_{\Phi^{a_1}}(z_1)\ V_{\Phi^{a_2}}(z_2)\
V_{\Phi^{a_3}}(z_3)\ V_{\Phi^{a_4}}(z_4)}\ .}
The vertex operators for the gauginos are given in \fieldsii, while the
vertex operators for the scalars and vectors are given in \Vertexop.
The four--point correlator in the integrand of \Vier\ is evaluated by performing
all possible Wick contractions. Some of the relevant correlators are
\eqn\Green{\eqalign{
\vev{S_\al(z_1)S_\bet(z_2)}&=\eps_{\al\bet}\ z_{12}^{-1/2}\ \ \ ,\ \ \
\vev{S_\al(z_1)S_{\dot\bet}(z_2)}=0\ ,\cr
\vev{S_\al(z_1)S_{\dot\beta}(z_2)\psi^\mu(z_3)}&=\fc{1}{\sqrt 2}\
\sigma^\mu_{\al\dot\bet}\ z_{13}^{-1/2}z_{23}^{-1/2}\ ,\cr
\vev{S_\al(z_1)S_\beta(z_2)\psi^\mu(z_3)}&=0\ ,}}
and:
\eqn\Ghost{
\vev{e^{-\h\phi(z_1)}e^{-\h\phi(z_2)}e^{-\h\phi(z_3)}e^{-\h\phi(z_4)}}=
(z_{12}\ z_{13}\ z_{14}\ z_{23}\ z_{24}\ z_{34})^{-1/4}\ .}
Again, $PSL(2,\IR)$ invariance on the disk allows to fix three vertex
positions according \choice,
which implies the ghost factor $\vev{c(z_1)c(z_2)c(z_3)}=-z_\infty^2$.
The remaining vertex position $z_4:=1/x$ takes arbitrary values inside
the integration domain $0<x<~1$ respecting the integration region
$z_1<\ldots<z_4$
in \Vier.

\noindent
{\it Four gauginos:}\br
For the four--gaugino amplitude involving four gauginos of the same chirality
in \Vier\ we compute the correlator
\eqn\startg{
\vev{V_{\lambda^{a_1,I}}^{(-1/2)}(z_1,u_1,k_1)\
V_{\lambda^{a_2,J}}^{(-1/2)}(z_2,u_2,k_2)\
V_{\lambda^{a_3,K}}^{(-1/2)}(z_3,u_3,k_3)\ V_{\lambda^{a_4,L}}^{(-1/2)}(z_4,u_4,k_4)}\ ,}
with the vertices \fieldsii. We need the following space--time fermion
correlator:
\eqn\split{\eqalign{
\vev{S_\al(z_1)S_\bet(z_2)S_\gamma(z_3)S_\delta(z_4)}&=
\fc{\epsilon_{\al\bet}\ \epsilon_{\gamma\delta}\ z_{13}z_{14}-
\epsilon_{\al\gamma}\ \epsilon_{\beta\delta}\ z_{12}z_{14}+
\epsilon_{\al\delta}\ \epsilon_{\beta\gamma}\ z_{12}z_{13}}
{(z_{12}\ z_{13}\ z_{14}\ z_{23}\ z_{24}\ z_{34})^{1/2}}\cr
&=\fc{\epsilon_{\al\bet}\ \epsilon_{\gamma\delta}\ z_{14}z_{23}-
\epsilon_{\al\delta}\ \epsilon_{\beta\gamma}\ z_{12}z_{34}}{
(z_{12}\ z_{13}\ z_{14}\ z_{23}\ z_{24}\ z_{34})^{1/2}}\ .}}
In the last step we have applied the Fierz identity
$\epsilon_{\al\gamma}\eps_{\beta\delta}=\eps_{\al\bet}\eps_{\gamma_\delta}+
\eps_{\al\delta}\eps_{\bet\gamma}$.
Furthermore, the correlator for the internal Ramond fields $\Si$ is:
\eqn\Split{
\vev{\Si^I(z_1)\Si^J(z_2)\Si^K(z_3)\Si^L(z_4)}=(z_{12}\ z_{13}\
z_{14}\ z_{23}\ z_{24}\ z_{34})^{-1/4}\ ,\ I\neq J\neq K\neq L\ .}
With the choice \choice, the correlators \Ghost, \split\ and \Split\ we arrive
at the partial amplitude of four gauginos:
\eqn\valentg{
A(\lambda^{I}\lambda^{J}\lambda^{K}\lambda^{L})=(2\ap)g_{Y\! M}^2\ \fc{\Gamma(s)\ \Gamma(u)}
{\Gamma(1+s+u)}\ \lf[\ u\ (u_1u_2)\ (u_3u_4)-s\ (u_1u_4)\ (u_2u_3)\ \ri]\ ,}
with the Mandelstam invariants $s=2\ap k_1k_2,\ t=2\ap k_1k_3$ and $u=2\ap k_1k_4$.
On the other hand, for the string amplitude with two gauginos of opposite chirality
\eqn\STartg{
\vev{V_{\lambda^{a_1,I}}^{(-1/2)}(z_1,u_1,k_1)\
V_{\lambda^{a_2,J}}^{(-1/2)}(z_2,u_2,k_2)\
V_{\ov\lambda^{a_3,K}}^{(-1/2)}(z_3,\ov u_3,k_3)\
V_{\ov\lambda^{a_4,L}}^{(-1/2)}(z_4,\ov u_4,k_4)}\ ,}
we need the correlator
\eqn\haveused{
\vev{S_\al(z_1)S_\bet(z_2)S^{\dot\gamma}(z_3)S^{\dot\delta}(z_4)}=
(z_{12}\ z_{34})^{-1/2}\ \epsilon_{\al\bet}\ \epsilon^{\dot\gamma\dot\delta}\ ,}
which may be derived by using the identity
$\sigma^\mu_{\al\dot\bet}\ov\sigma_\mu^{\dot\delta\gamma}=-
2\ \delta_\al^\gamma\ \delta_{\dot\beta}^{\dot\delta}$
and the correlator:
\eqn\Splitt{
\vev{\Si^I(z_1)\Si^J(z_2)\ov\Si^{K}(z_3)\ov\Si^{L}(z_4)}=
\lf(\fc{z_{13}\ z_{14}\ z_{23}\ z_{24}}{z_{12}\ z_{34}}\ri)^{1/4}\
\lf(-\fc{\delta^{IK}\delta^{JL}}{z_{13}\ z_{24}}+\fc{\delta^{IL}\
\delta^{JK}}{z_{14}\ z_{23}}\ri)\ .}
With the choice \choice, the correlators \Ghost, \haveused\ and \Splitt\
the partial amplitude becomes:
\eqn\valentgg{
A(\lambda^I\lambda^J\ov\lambda^K\ov\lambda^L)=-(2\ap)g_{Y\! M}^2\
(u_1u_2)\ (\ov u_3\ov u_4)\ \fc{\Gamma(s)\ \Gamma(u)}{\Gamma(1+s+u)}\
\lf(u\ \delta^{IK}\delta^{JL}+t\ \delta^{IL}\delta^{JK}\ri)\ .}

\noindent
{\it Two gauginos and two scalars:}\br
Now we compute the amplitude of two gauginos and two scalars by
calculating  in \Vier\ the correlator
\eqn\startsg{
\vev{V_{\lambda^{a_1,I}}^{(-1/2)}(z_1,u_1,k_1)\ V_{\phi^{a_2,j\mp}}^{(-1)}(z_2,k_2)\
V_{\ov\lambda^{a_3,K}}^{(-1/2)}(z_3,\ov u_3,k_3)\
V_{\phi^{a_4,l+}}^{(0)}(z_4,k_4)}\ .}
Here the index $j$ includes the two cases $j+$ and $j-$.
After deriving the set of correlators
\eqn\withcorr{\eqalign{
\vev{\Sigma^I(z_1)\ov\Sigma^I(z_3)\Psi^{j-}(z_2)\Psi^{j+}(z_4)}&=
\fc{1}{z_{24}\ z_{13}^{3/4}}\ \lf(\fc{z_{14}\ z_{32}}{z_{12}\ z_{34}}\ri)^{1/2}\ ,\cr
(a):\ \ \ (j,I)&\in\{(1,3),(1,4),\ (2,2),(2,4),\ (3,2),(3,3)\}\ ,\cr
\vev{\Sigma^I(z_1)\ov\Sigma^I(z_3)\Psi^{j-}(z_2)\Psi^{j+}(z_4)}&=
\fc{1}{z_{24}\ z_{13}^{3/4}}\
\lf(\fc{z_{12}\ z_{34}}{z_{14}\ z_{32}}\ri)^{1/2}\ ,\cr
(b):\ \ \ (j,I)&\in\{(1,1),(1,2)\ ,(2,1),(2,3), (3,1),(3,4)\}\ ,\cr
\vev{\Sigma^I(z_1)\ov\Sigma^K(z_3)\Psi^{j\mp}(z_2)\Psi^{l+}(z_4)}&=
z_{13}^{1/4}\lf(z_{12}\ z_{14}\ z_{32}\ z_{34}\ri)^{-1/2}\ ,\cr
(c):\ \ \ (j-,l,I,K)\in\{(1,2,3,2),&\ (1,3,4,2),\ (2,3,4,3),\ (2,1,2,3),\ (3,1,2,4),\
(3,2,3,4)\}\ ,\cr
\ \ \ (j+,l,I,K)\in\{(1,2,1,4),&\ (1,3,1,3),\ (2,3,1,2),\ (2,1,1,4),\ (3,1,1,3),\
(3,2,1,2)\}\ ,}}
with \Green\ the resulting partial amplitude becomes
\eqn\Valentggs{
A(\lambda^I\phi^{j\mp}\ov\lambda^K\phi^{l+})=(2\ap) g_{Y\! M}^2\
\fc{\Gamma(s)\ \Gamma(u)}{\Gamma(1+s+u)}\
k_{4\mu}\ (u^\al_1\sigma^\mu_{\al\dot\bet}\ov u_3^{\dot\bet})\times
\cases{
u\ ,&case $(a)$\ ,\cr
-s\ ,&case $(b)$\ ,\cr
t\ ,&case $(c)$\ ,}}
for the three cases in \withcorr.

\noindent
{\it Two gauginos, one vector and one scalar:}\br
Next, we compute the four--point amplitude of two gauginos, one vector and one
scalar. In \Vier\ we compute the correlator:
\eqn\startsgs{
\vev{V_{\lambda^{a_1,I}}^{(-1/2)}(z_1,u_1,k_1)\
V_{\lambda^{a_2,J}}^{(-1/2)}(z_2,u_2,k_2)\
V_{A^{a_3}}^{(-1)}(z_3,\xi_3,k_3)\ V_{\phi^{a_4,l+}}^{(0)}(z_4,k_4)}\ .}
For the space--time fermions  we need the correlator:
\eqn\Splitt{\eqalign{
\vev{S_\al(z_1)S_\bet(z_2)\psi^\mu(z_3)\psi^\nu(z_4)}&=
z_{34}^{-1}\ (z_{12}\ z_{13}\ z_{14}\ z_{23}\ z_{24})^{-1/2}\cr
&\hskip-1cm\times\lf(\epsilon_{\al\bet}\ \delta^{\mu\nu}\ z_{13}z_{14}+\h
\sigma^\mu_{\al\dot\gamma}\ \epsilon^{\dot\gamma\dot\delta}\
\sigma_{\bet\dot\delta}^\nu\ z_{12}z_{14}+\h
\sigma^\nu_{\al\dot\gamma}\ \epsilon^{\dot\gamma\dot\delta}\ \sigma^\mu_{\bet\dot\delta}
\ z_{12}z_{13}\ri)\cr
&=\h\ \fc{\epsilon_{\al\bet}\ \delta^{\mu\nu}\ (z_{13}z_{24}+z_{14}z_{23})}
{z_{34}\ (z_{12}\ z_{13}\ z_{14}\ z_{23}\ z_{24})^{1/2}}-\h\
\fc{\si^{\mu\nu}_{\al\bet}\ z_{12}^{1/2}}
{(z_{13}\ z_{14}\ z_{23}\ z_{24})^{1/2}}\ .}}
The last step follows from the relations
$\ov\si^{\mu\dot\al\al}=\eps^{\dot\al\dot\bet}\eps^{\al\bet}
\si^\mu_{\bet\dot\bet}$ and $\si^\nu_{\bet\dot\delta}\ov\si^{\mu\dot\delta\lambda}+
\si^\mu_{\bet\dot\delta}\ov\si^{\nu\dot\delta\lambda}=-2\delta^{\mu\nu}\delta^\lambda_\bet$.
The necessary correlator for the internal fields is:
\eqn\WWith{
\vev{\Si^I(z_1)\Si^J(z_2)\Psi^{l+}(z_4)}=z_{12}^{-1/4}
\lf(z_{14}z_{24}\ri)^{-1/2}\ \ \ ,\ \ \ (I,J,l+)\in\Ic\ ,}
with $\Ic=\{(1,2,1),(1,3,2),(1,4,3), (2,1,1),(3,1,2),(4,1,3)\}$.
With \Splitt\ and \WWith\ the partial amplitude of \Vier\ becomes
\eqn\Valentggss{
A(\lambda^I\lambda^JA\phi^{l+})=\sqrt 2\ap g_{Y\! M}^2\
\fc{\Gamma(s)\ \Gamma(u)}{\Gamma(1+s+u)}\
\lf[\ (u-t)\ (u_1u_2)(\xi_3k_4)-s\ (u_1^\al\si_{\al\bet}^{\mu\nu}u_2^\bet)
\ \xi_{3\mu} k_{4\nu}\ \ri]\ ,}
with $(I,J,l+)\in\Ic$.

\noindent
{\it Four scalars:}\br
Finally we compute the string $S$--matrix of four scalars.
With the vertex operators of \fieldsi\ in \Vier\ we compute the correlator:
\eqn\starts{
\vev{V_{\phi^{a_1,i-}}^{(0)}(z_1,k_1)\ V_{\phi^{a_2,j-}}^{(0)}(z_2,k_2)\
V_{\phi^{a_3,k+}}^{(-1)}(z_3,k_3)\ V_{\phi^{a_4,l+}}^{(-1)}(z_4,k_4)}\ .}
After some algebra and using \Split\
we find the following expression for the partial amplitude:
\eqn\valents{
A(\phi^{i-}\phi^{j-}\phi^{k+}\phi^{l+})=-2g_{Y\! M}^2\
\fc{\Gamma(s)\ \Gamma(u)}{\Gamma(1+s+u)}\
\lf(u\ \delta^{ik}\delta^{jl}+t\ \delta^{il}\delta^{jk}\ri)\ s\ .}

\subsec{Supersymmetry relations of string amplitudes}

By applying the results of Subsection 2.3 in this part we derive
relations between the string amplitudes we have computed in the previous Subsection.
Inspection of those amplitudes shows that they all
have the same prefactor encoding the $\ap$-dependence of the amplitude.
Indeed here we show that they all related through supersymmetry
transformations.
With \susytransi\ we may derive a SUSY relation between a four gaugino amplitude
and two amplitudes involving two gauginos and two scalars.

In analogy to \NPOINT\ we start with the contour integral
\eqn\NNPOINTs{
\oint\limits_{C_\infty}\fc{dw}{2\pi i}
\eta_M^\al\ \vev{V_\al^M(w)\ V_{\lambda^{a_1,I}}^{(-1/2)}(z_1,u_1,k_1)\
V_{\lambda^{a_2,J}}^{(-1/2)}(z_2,u_2,k_2)\
V_{\ov\lambda^{a_3,K}}^{(-1/2)}(z_3,\ov u_3,k_3)\
V_{\phi^{a_4,l+}}^{(0)}(z_4,k_4)}\ ,}
where $C_\infty$ is a closed contour in the complex plane encircling all four
vertex positions $z_1\ldots,z_4$. With the arguments of Subsection 2.3
the integral \NNPOINTs\ vanishes.
On the other hand, by analyticity in \NNPOINTs\ we may deform the contour
to the other four vertex operators with the SUSY operator acting
on one vertex operator, respectively (c.f. Eq. \Susytrans):
\eqn\Action{\eqalign{
&[Q^M(\eta_M), V_{\lambda^{a_1,I}}^{(-1/2)}(z_1,u_1,k_1)]=\vev{\eta_M u_1}\
V_{\phi^{a_1,i\mp}}^{(-1)}(z_1,u_1,k_1)\ ,\cr
&[Q^M(\eta_M), V_{\lambda^{a_2,J}}^{(-1/2)}(z_2,u_2,k_2)]=
\vev{\eta_M u_2}\ V_{\phi^{a_2,j\mp}}^{(-1)}(z_2,k_2)\ ,\cr
&[Q^M(\eta_M), V_{\ov\lambda^{a_3,K}}^{(-1/2)}(z_3,\ov u_3,k_3)]=\fc{1}{\sqrt
2}\ \delta^{MK}\ V_{A^{a_3}}^{(-1)}(z_3,\xi_3)\ \ \ ,\ \ \
\xi_3^\mu=\eta_M^\al\sigma^\mu_{\al\dot\gamma}\ov u_3^{\dot\gamma}\ ,\cr
&[Q^M(\eta_M), V_{\phi^{a_4,l+}}^{(0)}(z_4,k_4)]=
V_{\ov\lambda^{a_4,L}}^{(-1/2)}(z_4,\ov u_4,k_4)\ \ \ ,\ \ \
\ov u_{4\dot\beta}=k_{4\rho}\ \eta_M^\al\sigma_{\al\dot\bet}^\rho\ .}}
Inserting these results into \WARD\ gives a sum of the following four correlators
\eqn\threecontri{\eqalign{
&\vev{\eta_M u_1}\
\vev{V_{\phi^{a_1,i\mp}}^{(-1)}(z_1,u_1,k_1)\
V_{\lambda^{a_2,J}}^{(-1/2)}(z_2,u_2,k_2)\
V_{\ov\lambda^{a_3,K}}^{(-1/2)}(z_3,\ov u_3,k_3)\ V_{\phi^{a_4,l+}}^{(0)}(z_4,k_4)}\cr
&=-2\ap g_{Y\! M}^2\ \fc{\Gamma(s)\ \Gamma(u)}{\Gamma(1+s+u)}\
k_{4\mu}\ \ \vev{\eta_M u_1}\
(u^\al_2\sigma^\mu_{\al\dot\bet}\ov u_3^{\dot\bet})\
\lf(u\ \delta^{IK}\delta^{JL}+t\ \delta^{IL}\delta^{JK}\ri)\ ,}}
where we have used \Valentggs. Similarly, the second correlator becomes:
\eqn\threecontrii{\eqalign{
&\vev{\eta_M u_2}\
\vev{V_{\lambda^{a_1,I}}^{(-1/2)}(z_1,u_1,k_1)\
V_{\phi^{a_2,j\mp}}^{(-1)}(z_2,k_2)\
V_{\ov\lambda^{a_3,K}}^{(-1/2)}(z_3,\ov u_3,k_3)\
V_{\phi^{a_4,l+}}^{(0)}(z_4,k_4)}\cr
&=2\ap g_{Y\! M}^2\ \fc{\Gamma(s)\ \Gamma(u)}{\Gamma(1+s+u)}\
k_{4\mu}\ \vev{\eta_M u_2}\
(u^\al_1\sigma^\mu_{\al\dot\bet}\ov u_3^{\dot\bet})\
\lf(u\ \delta^{IK}\delta^{JL}+t\ \delta^{IL}\delta^{JK}\ri)\ .}}
With \Valentggss\ the third correlator gives:
\eqn\threecontriii{\eqalign{
&\fc{1}{\sqrt 2}\ \delta^{MK}\ \vev{V_{\lambda^{a_1,I}}^{(-1/2)}(z_1,u_1,k_1)\
V_{\lambda^{a_2,J}}^{(-1/2)}(z_2,u_2,k_2)\ V_{A^{a_3}}^{(-1)}(z_3,\xi_3)\
V_{\phi^{a_4,l+}}^{(0)}(z_4,k_4)}\cr
&=\ap g_{Y\! M}^2\ \fc{\Gamma(s)\ \Gamma(u)}{\Gamma(1+s+u)}\
\lf[(u-t)\vev{u_1u_2}(\xi_3k_4)-s(u_1^\al\si_{\al\bet}^{\mu\nu}u_2^\bet)
\ \xi_3^\mu k_4^\nu\ri]\delta^{MK}\ ,\
\xi_3^\mu=\eta_M^\al\sigma^\mu_{\al\dot\gamma}\ov  u_3^{\dot\gamma}\ ,\cr
&=2\ap g_{Y\! M}^2\ \fc{\Gamma(s)\ \Gamma(u)}{\Gamma(1+s+u)}\
k_{4\mu}\lf[u\vev{u_1u_2}(\eta_M^\al\si_{\al\dot\bet}^\mu\ov u_3^{\dot\bet})
-s\vev{\eta_Mu_2}
(u_1^\al\si^\mu_{\al\dot\bet}\ov u_3^{\dot\bet})\ri]\delta^{MK},\
(I,J,l+)\in\Ic\ .}}
Finally, with \valentgg\ the last correlator becomes:
\eqn\threecontriv{\eqalign{
&\vev{V_{\lambda^{a_1,I}}^{(-1/2)}(z_1,u_1,k_1)\
V_{\lambda^{a_2,J}}^{(-1/2)}(z_2,u_2,k_2)\
V_{\ov\lambda^{a_3,K}}^{(-1/2)}(z_3,\ov u_3,k_3)\
V_{\ov\lambda^{a_4,L}}^{(-1/2)}(z_4,\ov u_4,k_4)}\cr
&=-2\ap g_{Y\! M}^2\ \fc{\Gamma(s)\ \Gamma(u)}{\Gamma(1+s+u)}\ \vev{u_1u_2}\ [u_3u_4]\
\lf(u\ \delta^{IK}\delta^{JL}+t\ \delta^{IL}\delta^{JK}\ri)\ \ \ ,\ \ \
\ov u_{4\dot\beta}=k_{4\rho}\ \eta_M^\al\sigma_{\al\dot\bet}^\rho\ .}}
For $M\neq K$ the third correlator \threecontriii\ vanishes. In that case
summing up \threecontri, \threecontrii\ and \threecontriv\ gives
$$\eqalign{
2\ap g_{Y\! M}^2\ &\fc{\Gamma(s)\ \Gamma(u)}{\Gamma(1+s+u)}\
\lf(u\ \delta^{IK}\delta^{JL}+t\ \delta^{IL}\delta^{JK}\ri)\cr
&\times k_{4\mu}\ \lf[-\vev{\eta_M u_1}\ (u_2^\al\si_{\al\dot\bet}^\mu\ov u_3^{\dot\bet})
+\vev{\eta_M u_2}\ (u^\al_1\sigma^\mu_{\al\dot\bet}\ov u_3^{\dot\bet})
-\vev{u_1u_2}\ (\eta_M^\al\si_{\al\dot\bet}^\mu\ov u_3^{\dot\bet})\ \ri]\ ,}$$
which vanishes due to the Fierz relation:
\eqn\fierz{
\vev{\eta_M u_2}\ (u_1^\al\si^\mu_{\al\dot\be}\ov u_3^{\dot\bet})=
\vev{\eta_M u_1}\ (u_2^\al\si_{\al\dot\bet}^\mu\ov u_3^{\dot\bet})
+(\eta_M^\al\si_{\al\dot\bet}^\mu\ov u_3^{\dot\bet})\ \vev{u_1u_2}\ .}
To conclude, to all orders in $\ap$ we have proven the following SUSY identity
\eqn\Derive{
\vev{\eta_M u_1}\ A(\phi^{i\mp}\lambda^{J}\ov\lambda^K\phi^{l+})+\vev{\eta_M u_2}\
A(\lambda^I\phi^{j\mp}\ov\lambda^K\phi^{l+})
+\lf.A(\lambda^I\lambda^{J}\ov\lambda^K\ov\lambda^L)=0\ \ri|_{\ov u_{4\dot\bet}=
k_{4\rho}\ \eta_M^\al\si_{\al\dot\bet}^\rho}\ ,}
for an arbitrary spinor $\eta_M$ and $M\neq K$.
Table 2 shows the combination of indices, for which all the three correlators
\threecontri, \threecontrii\ and \threecontriv\ are non--vanishing, but
\threecontrii\ vanishing due to $M\neq K$.
\vskip0.3cm
{\hskip-5.5cm\vbox{\ninepoint{
$$
\vbox{\offinterlineskip\tabskip=0pt
\halign{\strut\vrule#
%%%%%%%%%%%%%%%%%%
&~$#$~\hfil
&\vrule#
&~$#$~\hfil
&~$#$~\hfil
&\vrule#&\vrule#
\cr
%%%%%%%%%%%%%%%%%%
\noalign{\hrule}
&
M
&&
(I,J,K,L,\ i,\ j,\ l)
&&
\cr
%%%%%%%%%%%%%%%%%%
\noalign{\hrule}
&
1
&&
(2,2,2,2,1-,1-,1+)
&&
\cr
%%%%%%%%%%%%%%%%%%
\noalign{\hrule}
&
1
&&
(2,3,3,2,1-,2-,1+)
&&
\cr
%%%%%%%%%%%%%%%%%%
%%%%%%%%%%%%%%%%%%
\noalign{\hrule}
&
1
&&
(2,3,2,3,1-,2-,2+)
&&
\cr
%%%%%%%%%%%%%%%%%%
\noalign{\hrule}
&
1
&&
(2,4,4,2,1-,3-,1+)
&&
\cr
%%%%%%%%%%%%%%%%%%
\noalign{\hrule}
&
1
&&
(2,4,2,4,1-,3-,3+)
&&
\cr
%%%%%%%%%%%%%%%%%%
\noalign{\hrule}
&
1
&&
(3,2,3,2,2-,1-,1+)
&&
\cr
%%%%%%%%%%%%%%%%%%
\noalign{\hrule}
&
1
&&
(3,3,3,3,2-,2-,2+)
&&
\cr
%%%%%%%%%%%%%%%%%%
\noalign{\hrule}
&
1
&&
(3,2,2,3,2-,1-,2+)
&&
\cr
%%%%%%%%%%%%%%%%%%
\noalign{\hrule}
&
1
&&
(3,4,4,3,2-,3-,2+)
&&
\cr
%%%%%%%%%%%%%%%%%%
%%%%%%%%%%%%%%%%%%
\noalign{\hrule}}}$$
\vskip-2pt
{\hskip4.5cm\noindent{\bf Table 2:}
{\sl Index structure for the SUSY relation \Derive}}
}}}
\vskip-5.9cm
{\hskip-1cm\vbox{\ninepoint{
$$
\vbox{\offinterlineskip\tabskip=0pt
\halign{\strut\vrule#
%%%%%%%%%%%%%%%%%%
&~$#$~\hfil
&\vrule#
&~$#$~\hfil
&~$#$~\hfil
&\vrule#&\vrule#
\cr
%%%%%%%%%%%%%%%%%%
\noalign{\hrule}
&
M
&&
(I,J,K,L,\ i,\ j,\ l)
&&
\cr
%%%%%%%%%%%%%%%%%%
\noalign{\hrule}
&
1
&&
(3,4,3,4,2-,3-,3+)
&&
\cr
%%%%%%%%%%%%%%%%%%
\noalign{\hrule}
&
1
&&
(4,3,3,4,3-,2-,3+)
&&
\cr
%%%%%%%%%%%%%%%
\noalign{\hrule}
&
1
&&
(4,4,4,4,3-,3-,3+)
&&
\cr
%%%%%%%%%%%%%%%%%%
%%%%%%%%%%%%%%%%%%
\noalign{\hrule}
&
1
&&
(4,2,4,2,3-,1-,1+)
&&
\cr
%%%%%%%%%%%%%%%%%%
\noalign{\hrule}
&
1
&&
(4,2,2,4,3-,1-,3+)
&&
\cr
%%%%%%%%%%%%%%%%%%
\noalign{\hrule}
&
1
&&
(4,3,4,3,3-,2-,2+)
&&
\cr
%%%%%%%%%%%%%%%%%%
\noalign{\hrule}
&
2
&&
(1,1,1,1,1-,1-,1+)
&&
\cr
%%%%%%%%%%%%%%%%%%
\noalign{\hrule}
&
3
&&
(1,1,1,1,2-,2-,2+)
&&
\cr
%%%%%%%%%%%%%%%%%%
\noalign{\hrule}
&
4
&&
(1,1,1,1,3-,3-,3+)
&&
\cr
%%%%%%%%%%%%%%%%%%
%%%%%%%%%%%%%%%%%%
\noalign{\hrule}}}$$
}}}
\vskip1cm
\vskip-6.77cm
{\hskip4cm\vbox{\ninepoint{
$$
\vbox{\offinterlineskip\tabskip=0pt
\halign{\strut\vrule#
%%%%%%%%%%%%%%%%%%
&~$#$~\hfil
&\vrule#
&~$#$~\hfil
&~$#$~\hfil
&\vrule#&\vrule#
\cr
%%%%%%%%%%%%%%%%%%
\noalign{\hrule}
&
M
&&
(I,J,K,L,\ i,\ j,\ l)
&&
\cr
%%%%%%%%%%%%%%%%%%
\noalign{\hrule}
&
2
&&
(1,3,3,1,1-,3+,1+)
&&
\cr
%%%%%%%%%%%%%%%
\noalign{\hrule}
&
2
&&
(1,4,4,1,1-,2+,1+)
&&
\cr
%%%%%%%%%%%%%%%%%%
\noalign{\hrule}
&
2
&&
(3,1,3,1,3+,1-,1+)
&&
\cr
%%%%%%%%%%%%%%%%%%
%%%%%%%%%%%%%%%%%%
\noalign{\hrule}
&
2
&&
(4,1,4,1,2+,1-,1+)
&&
\cr
%%%%%%%%%%%%%%%%%%
\noalign{\hrule}
&
3
&&
(1,2,2,1,2-,3+,2+)
&&
\cr
%%%%%%%%%%%%%%%%%%
\noalign{\hrule}
&
3
&&
(1,4,4,1,2-,1+,2+)
&&
\cr
%%%%%%%%%%%%%%%%%%
\noalign{\hrule}
&
3
&&
(2,1,2,1,3+,2-,2+)
&&
\cr
%%%%%%%%%%%%%%%%%%
\noalign{\hrule}
&
3
&&
(4,1,4,1,1+,2-,2+)
&&
\cr
%%%%%%%%%%%%%%%%%%
\noalign{\hrule}
&
4
&&
(1,2,2,1,3-,2+,3+)
&&
\cr
%%%%%%%%%%%%%%%%%%
\noalign{\hrule}
&
4
&&
(1,3,3,1,3-,1+,3+)
&&
\cr
%%%%%%%%%%%%%%%%%%
\noalign{\hrule}
&
4
&&
(2,1,2,1,2+,3-,3+)
&&
\cr
%%%%%%%%%%%%%%%%%%
\noalign{\hrule}
&
4
&&
(3,1,3,1,1+,3-,3+)
&&
\cr
%%%%%%%%%%%%%%%%%%
%%%%%%%%%%%%%%%%%%
\noalign{\hrule}}}$$
}}}
%\vskip0.25cm
\noindent
On the other hand, in the case $M=K$ the third correlator \threecontriii\ contributes
in \WARD. However, in that case one of the three correlators
\threecontri, \threecontrii\ or \threecontriv\ vanishes because the
corresponding SUSY transformation \Action\ gives a vanishing contribution.
In either case the three non--vanishing residua sum up to zero due to \fierz.
E.g. for $M=K=2, I=1, J=2$ and $l=1$ the SUSY transformations \Action\ yield
$i=1, L=1$ but vanishing SUSY transformation for the variation of
$V_{\lambda^{a_2,J}}$.
Hence the total contribution of residua to \NNPOINTs\ is:
$$\eqalign{
2\ap g_{Y\! M}^2\ \fc{\Gamma(s)\ \Gamma(u)}{\Gamma(1+s+u)}\
&k_{4\mu}\ \lf[-s\ \vev{\eta_M u_1}\ (u_2^\al\si_{\al\dot\bet}^\mu\ov u_3^{\dot\bet})
-t\ \vev{u_1u_2}\ (\eta_M^\al\si_{\al\dot\bet}^\mu\ov u_3^{\dot\bet})\ri.\cr
&-\lf. u\ \vev{u_1u_2}\ (\eta_M^\al\si_{\al\dot\bet}^\mu\ov u_3^{\dot\bet})
+s\ \vev{\eta_Mu_2}\
(u_1^\al\si^\mu_{\al\dot\bet}\ov u_3^{\dot\bet})\ \ri]\ ,}$$
which vanishes according to \fierz.
Eventually \WARD\ gives rise to the following identity:
\eqn\DDerive{
\vev{\eta_2 u_1}\ A(\phi^{1+}\lambda^{2}\ov\lambda^2\phi^{1+})+
\fc{1}{\sqrt 2}\ A(\lambda^1\lambda^2A\phi^{1+})
+\lf.A(\lambda^1\lambda^2\ov\lambda^2\ov\lambda^1)=0\ \ri|_{\ov u_{4\dot\bet}=
k_{4\rho}\ \eta_2^\al\si_{\al\dot\bet}^\rho\atop \xi_3^\mu=\eta_M^\al
\sigma^\mu_{\al\dot\gamma}\ov  u_3^{\dot\gamma}}\ .}

\newsec{Concluding Remarks}

The main result of this paper is the extension of the well-known SUSY
Ward identities relating the scattering amplitudes of particles with different
spin but belonging to the same supermultiplet, to type I open superstring
theory
at the disk level. {}For arbitrary compactifications, the form of such
relations
remains exactly the same as in the low-energy effective field theory
describing the
$\alpha'\to 0$ limit.
Although we studied explicitly only the case of gauge supermultiplets it is
clear
that our results extend to ``matter'' supermultiplets.

Here, we focussed on one particular application of SUSY relations: to the computations of $N$-gluon MHV amplitudes. In this case, four gluons can be replaced by scalars, leading to significant simplifications. By using this method, we were able to reproduce all known results for $N\le 6$ and to derive a compact expression for the seven-gluon MHV amplitude.
It is very interesting that the use of supersymmetry dictates certain choice
of
$(N{-}3)!$ elements of the basis of boundary integrals over the vertex positions, such that MHV amplitudes can be represented as linear combinations of these basis functions weighted by very simple twistor-like coefficients. It seems that such a representation is more natural than the factorized form discussed previously in \doubref\STi\STii. We believe that it is also more suitable for studying the recursive structure for arbitrary $N$.

There are several extensions and applications of our results that deserve further studies. In particular, it would be interesting to see if SUSY relations 
lead to a novel representation of superstring amplitudes also for 
non-MHV configurations (that appear for $N\ge 6$) \STiii. 
Furthermore, a generalization of SUSY relations to world-sheets with string loops would be useful for understanding how supersymmetry is realized at the level of quantum-corrected, or perhaps even exact, scattering amplitudes.
\vskip2cm
\centerline{\noindent{\bf Acknowledgments} }
This work  is supported in part by the European Commission under
Project MRTN-CT-2004-005104.
The research of T.R.T.\ is supported in part by the U.S.
National Science Foundation Grants PHY-0242834 and PHY-0600304.
T.R.T.\ is grateful to Dieter L\"{u}st,
to Arnold Sommerfeld Center for Theoretical Physics at Ludwig--Maximilians
Universit\"at, and to Max-Planck-Institut f\"ur Physik
in M\"unchen, for their kind hospitality.
St.St. is indebted to CERN and Northeastern University in Boston for warm
hospitality and financial support.
He would like to also thank the Galileo Galilei Institute for Theoretical
Physics in Firenze for
hospitality and INFN for partial support during completion of this work.
Any opinions, findings, and conclusions or
recommendations expressed in this material are those of the authors and do
not necessarily reflect the views of the National Science Foundation.

\break
\appendix\appA{Operator product expansions and extended supersymmetry}

In this appendix we present the $N$ holomorphic spacetime supersymmetry currents
and their operator product expansions. We adapt to the notation and spinor
algebra of the book of Wess and Bagger. In particular, spinor indices are raised and
lowered with the anti--symmetric tensors $\eps_{\al\bet}$ and $\eps^{\dot\al\dot\bet}$.
Besides spinor products are defined to be
$\chi\eta=\chi^\al\eps_{\al\bet}\eta^\bet$ ($\ov\chi\ov\eta=\ov\chi_{\dot\al}
\eps^{\dot\al\dot\bet}\ov\eta_{\dot\bet}$) for some spinors $\chi,\eta$ ($\ov\chi,\ov\eta$).
The $D=4$ supersymmetry currents in the $(-1/2)$--ghost picture have already been given
in \Susycurrents.
The internal Ramond fields $\Sigma^I$, which belong to an
internal superconformal field theory with $c=9$, have conformal dimensions $3/8$,
while the space--time spin fields $S_\al$ have conformal dimensions $1/4$
w.r.t. the holomorphic stress tensor $T(z)$.
The supersymmetry currents in the $(+1/2)$--ghost picture take the form:
\eqn\susychargesi{
V_\al^I(z)=i\ap^{-1/4}\ \fc{e^{\h\phi}}{(2\ap)^{1/2}}\ \lf(\ \fc{1}{\sqrt 2}\
\sigma^\mu_{\al\dot{\beta}}\ S^{\dot\beta}\ \p X_\mu\
\Sigma^I+S_\al\ \tilde\Si^I\ \ri)\ .}
The dimension $11/8$ internal conformal field $\tilde\Si^I$ appears in
the operator product expansion (OPE) of the field $\Si^I$ with the internal
super current: $\Si^I(z)\ T^{\rm
int}_F(w)=i(2\ap)^{-1/2}(z-w)^{-1/2}\ \tilde\Si^I(w)+~\ldots$. The full super
current is $T_F=i(2\ap)^{-1/2} \p X^\mu\psi_\mu+T^{\rm int}_F$.
The supersymmetry currents $V_\al^I$ have conformal dimension one.
The relevant OPEs for the spin fields are in $D=4$ \banksi:
\eqn\OPE{\eqalign{
e^{q_1\phi(z)}\ e^{q_2\phi(w)}&=(z-w)^{-q_1q_2}\
e^{(q_1+q_2)\phi(w)}+\ldots\ ,\cr
S_\al(z)S_{\dot\beta}(w)&=\fc{1}{\sqrt2}(z-w)^0\sigma_{\al\dot\beta}^\mu
\psi_\mu(w)+\ldots,\
S^{\dot\al}(z)S^{\beta}(w)=\fc{1}{\sqrt2}(z-w)^0\ov\sigma^{\mu\dot\al\beta}
\psi_\mu(w)+\ldots,\cr
S_\al(z)S_{\beta}(w)&=(z-w)^{-1/2}\ \epsilon_{\al\beta}+\ldots\ ,\
S^{\dot\al}(z)S^{\dot\beta}(w)=-(z-w)^{-1/2}\
\epsilon^{\dot\al\dot\beta}+\ldots\ ,\cr
\psi^\mu(z)S_\al(w)&=\fc{1}{\sqrt2}(z-w)^{-1/2} \sigma^\mu_{\al\dot\beta}\
S^{\dot\beta}(w)-\fc{1}{\sqrt2}(z-w)^{1/2}\psi^\mu(w)\psi_\nu(w)
\sigma^\nu_{\al\dot\bet}\ S^{\dot\bet}(w)+\ldots,\cr
\psi^\mu(z)S^{\dot\al}(w)\ &=\fc{1}{\sqrt2}(z-w)^{-1/2} \ov\sigma^{\mu\dot\al\beta}\
S_\beta(w)-\fc{1}{\sqrt2}(z-w)^{1/2}\psi^\mu(w)\psi_\nu(w)
\ov\sigma^{\nu\dot\al\bet}S_{\bet}(w)+\ldots\;.}}
Furthermore we need the following two OPEs:
\eqn\NEEDOPE{\eqalign{
S_\al(z)\ \psi^\mu(w)\psi^\nu(w)&=\h\ (z-w)^{-1}\
(\sigma^{\mu\nu})_\al^{\ \ \beta}\ S_\bet(z)+\ldots\ ,\cr
S^{\dot\al}(z)\ \psi^\mu(w)\psi^\nu(w)&=\h\ (z-w)^{-1}\
(\ov\sigma^{\mu\nu})^{\dot\al}_{\ \ \dot\beta}\ S^{\dot\bet}(z)+\ldots\ .}}
It is convenient to represent\foot{In these bosonized expressions we neglect
cocycle factors which are required to obtain $SO(4)$ covariant correlation
functions \KLLSW.} the spin fields $S_\al,\ S_{\dot\al}$ as
exponentials
of two free bosons $H^1, H^2$: $S_\al=e^{i\al H},\
S_{\dot\al}=e^{i\dot\al H}$, with $\al=(\pm\h,\pm\h)$ and
$\dot\al=(\pm\h,\mp\h)$, respectively. Their corresponding OPEs are:
\eqn\corrOPE{
e^{is_1H^i(z)}\ e^{is_2H^j(w)}=\delta^{ij}\ (z-w)^{s_1s_2}\
e^{i(s_1+s_2)H^i(w)}+\Oc((z-w)^{s_1s_2+1})\ .}
The OPEs of two supercharges \susy
\eqn\SUS{\eqalign{
\Qc_\al^I(z)\ \ov \Qc_{\dot\beta}^J(w)&=\fc{\ap^{-1/2}}{\sqrt2(2\pi i)^2}\
\oint\oint (z-w)^{-1/4}\
\fc{i\p X_\mu}{(2\ap)^{1/2}}\ \sigma^\mu_{\alpha\dot\beta}\ \Sigma^I(z)\
\ov\Sigma^J(w)+\ldots \ ,\cr
\Qc_\al^I(z)\ \Qc_{\beta}^J(w)&=\fc{\ap^{-1/2}}{(2\pi i)^2}\ \oint\oint e^{-\phi(w)}\
(z-w)^{-3/4}\ \epsilon_{\al\bet}\
\Sigma^I(z)\ \Sigma^{J}(w)+\ldots}}
and the OPEs  \ope\ of the internal Ramond fields ~$\Sigma^I$
reproduce the spacetime supersymmetry algebra
\eqn\spacetimesusy{
\{\Qc^I_\al,\ov \Qc_{\dot\bet}^J\}=\delta^{IJ}\ \sigma_{\al\dot\bet}^\mu\ P_\mu\ \ \
,\ \ \ \{\Qc^I_\al,\Qc_\bet^J\}=\epsilon_{\al\bet}\ Z^{IJ}\ ,}
with $P_\mu=i\oint \limits_{w=z} \fc{dw}{2\pi i}\ \fc{\p X_\mu}{2\ap}$
and the central charges
$Z^{IJ}=\ap^{-1/2}\oint\limits_{w=z} \fc{dw}{2\pi i}\ e^{-\phi(w)}\ \psi^{IJ}(w)$
(in the $(-1)$--ghost picture) of the extended supersymmetry algebra.
The latter correspond to the compactified fields $\p Z^i$.

\appendix\appB{Material for six--point function}

In this Appendix we present some additional material for the six--point
function \Becomes, which has been computed in Subsection 3.2.
In addition to the integrals \Functions\ the remaining seven integrals $K_i$ determining
the complete seven--point function \Becomes\ are:
\eqn\AddFunctions{\eqalign{
K_7&=\int\limits_0^1 dx\int\limits_0^1 dy\ \int\limits_0^1 dz\
\fc{s_{12}\ (1-x)}{x(1-z)(1-xy)(1-yz)}\ \Ic(x,y,z)\ ,\cr
K_8&=-\int\limits_0^1 dx\int\limits_0^1 dy\ \int\limits_0^1 dz\
\fc{s_{12}\ (1-x)}{x(1-y)(1-z)(1-xyz)}\ \Ic(x,y,z)\ ,\cr
K_9&=-\int\limits_0^1 dx\int\limits_0^1 dy\ \int\limits_0^1 dz\
\lf(1-s_{13}+\fc{s_{13}}{x}\ri)\ \fc{\Ic(x,y,z)}{(1-z)(1-yz)}\ ,\cr
K_{10}&=\int\limits_0^1 dx\int\limits_0^1 dy\ \int\limits_0^1 dz\
\lf(1-s_{13}+\fc{s_{13}}{x}\ri)\ \fc{\Ic(x,y,z)}{(1-y)(1-z)}\ ,\cr
K_{11}&=-\int\limits_0^1 dx\int\limits_0^1 dy\ \int\limits_0^1 dz\
\lf(\fc{1-s_{23}}{x}+s_{23}\ri)\ \fc{\Ic(x,y,z)}{xy(1-z)(1-yz)}\ ,\cr
K_{12}&=\int\limits_0^1 dx\int\limits_0^1 dy\ \int\limits_0^1 dz\
\lf(\fc{1-s_{23}}{x}+s_{23}\ri)\ \fc{\Ic(x,y,z)}{xyz(1-y)(1-z)}\ ,}}
$$\eqalign{K_{13}&=-\int\limits_0^1 dx\int\limits_0^1 dy\ \int\limits_0^1 dz\
\lf[(1-s_{13})(1-s_{24})+\fc{(1-s_{14})(1-s_{23})}{x^2}\ri.\cr
&\hskip4cm\lf.
+\fc{s_{12}s_{34}-s_{14}s_{23}-s_{13}s_{24}}{x}\ri]\ \fc{\Ic(x,y,z)}{y(1-z)^2}\ .}$$
The six--point amplitude \Becomes\ is completely specified by the basis of six functions
$\{K_1,\ldots,K_6\}$, introduced in \Functions.
Therefore the above integrals \AddFunctions\ can be expressed in terms of this basis:
\eqn\RelK{\eqalign{
\sss{(s_4+s_5-t_1)t_1K_7}&=\sss{(s_1+s_6-t_3) (s_3+s_4-t_1-t_3)
K_1+(s_1+s_4-t_1-t_3) (s_3+s_6-t_2-t_3)K_2+(s_5+s_6-t_2)}\cr 
&\sss{\times (s_1+s_6-t_3)K_3-(s_5+s_6-t_2)(s_1+s_4-t_1-t_3)K_4+s_6
   (s_3+s_6-t_2-t_3)K_5-s_6(s_3+s_4-t_1-t_3)K_6\ ,}\cr
\sss{s_4t_1\ K_8}&=\sss{-(s_3+s_4-t_3) (s_1+s_6-t_3)K_1-(s_1+s_4-t_3)(s_3+s_6-t_2-t_3) 
K_2-(s_5+s_6-t_2)}\cr
&\sss{\times (s_1+s_6-t_3)K_3
+(s_5+s_6-t_2) (s_1+s_4-t_3) K_4-s_6 (s_3+s_6-t_2-t_3) K_5+s_6 (s_3+s_4-t_3) K_6\ ,}\cr
\sss{(s_4+s_5-t_1) t_1 K_9}&=\sss{(s_3+s_4-t_3) (s_1+s_6-t_3)K_1+(s_1+s_4-t_1-t_3)
(s_3+s_6+t_1-t_2-t_3)K_2+(s_5+s_6-t_2)}\cr
&\sss{\times (s_1+s_6-t_3) K_3-(s_5+s_6-t_2) (s_1+s_4-t_1-t_3) K_4+s_6 
(s_3+s_6+t_1-t_2-t_3) K_5-s_6 (s_3+s_4-t_3) K_6\ ,}\cr
\sss{s_4t_1 K_{10}}&=\sss{-(s_3+s_4-t_3) (s_1+s_6-t_1-t_3)K_1-
(s_1+s_4-t_1-t_3) (s_3+s_6-t_2-t_3)K_2-(s_5+s_6-t_2) }\cr
&\sss{\times(s_1+s_6-t_1-t_3) K_3+
(s_5+s_6-t_2) (s_1+s_4-t_1-t_3)K_4-s_6(s_3+s_6-t_2-t_3) K_5+s_6(s_3+s_4-t_3)K_6}\ ,\cr
\sss{(s_4+s_5-t_1) t_1K_{11}}&=\sss{(s_3+s_4-t_3) (s_1+s_6-t_3)K_1+
(s_1+s_4-t_1-t_3) (s_3+s_6-t_2-t_3) K_2+(s_5+s_6-t_1-t_2)}\cr 
&\sss{\times (s_1+s_6-t_3)K_3-(s_5+s_6-t_1-t_2) (s_1+s_4-t_1-t_3) K_4+
s_6 (s_3+s_6-t_2-t_3) K_5-s_6 (s_3+s_4-t_3) K_6\ ,}\cr
\sss{s_4t_1 K_{12}}&=\sss{-(s_3+s_4-t_3) (s_1+s_6-t_3)K_1-(s_1+s_4-t_1-t_3) 
(s_3+s_6-t_2-t_3) K_2-(s_5+s_6-t_2) (s_1+s_6-t_3) K_3}\cr
&\sss{+(s_5+s_6-t_2) (s_1+s_4-t_1-t_3) K_4-(s_6+t_1)
(s_3+s_6-t_2-t_3) K_5+(s_6+t_1) (s_3+s_4-t_3) K_6\ ,}\cr
\sss{t_1 K_{13}}&=\sss{-(s_3+s_4-t_3) (s_1+s_6-t_3) K_1
-(s_1+s_4-t_1-t_3)(s_3+s_6-t_2-t_3) K_2-(s_5+s_6-t_2)}\cr
&\sss{\times (s_1+s_6-t_3) K_3+(s_5+s_6-t_2) (s_1+s_4-t_1-t_3) K_4-s_6 
(s_3+s_6-t_2-t_3) K_5+s_6 (s_3+s_4-t_3) K_6\ .}}}
In \Read\ we have expressed the function $K_3$ w.r.t. to the basis
$\{F_1,F_2,F_3,F_4,F_5,F_6\}$, introduced in \Functs. Similarly the remaining five
basis elements are expressed in terms of the basis $\{F_1,F_2,F_3,F_4,F_5,F_6\}$:
\eqn\RRead{\eqalign{
\sss{s_2\ s_5\ K_1}&=\sss{s_2 s_6 t_2 \ F_1+s_6 [-s_1 s_2+s_2 s_5-(s_4+s_5-t_1)
(s_5-t_2)]\ F_2+(s_4+s_5-t_1) (s_5-t_2) (-s_1-s_3+s_5+t_3)\ F_6}\cr
&\sss{+\lf\{-s_1^2 s_2+s_1s_2 (-s_4+s_5+t_1-t_2+t_3)-(s_5-t_2) [s_2 \
(s_5+t_3)-(s_4+s_5-t_1) (s_3-s_5+s_6+t_1-t_3)]\ri\} \ F_3}\cr
&\sss{-(s_4+s_5-t_1) [s_1 s_2+(s_5-t_2) (-s_3+s_5-t_1+t_3)]\ F_4}\cr
&\sss{+\lf\{s_1^2 \
s_2-(s_4+s_5-t_1) (s_3-s_5+t_1) (s_5-t_2)+s_1 [(s_4+s_5-t_1) (s_5-t_2)+s_2 \
(s_4-t_1-t_3)]\ri\} \ F_5}\cr}}
%%%%%%%%%%%%%%%%%%%%%%%%%%%%%%%%%%%%%%%%%%%%%%%%%%
$$\eqalign{\sss{s_2\ s_5\ K_2}&=\sss{-s_2 s_6 t_2 \ F_1+s_6 [s_1
s_2+(s_4+s_5-t_1) (s_5-t_2)] \
F_2+(s_4+s_5-t_1) (s_5-t_2) (s_1+s_3-s_5-t_3) \ F_6}\cr
&\sss{+\lf\{s_1^2 s_2+s_5^3-s_3 \
(s_4+s_5-t_1) (s_5-t_2)-t_1 (s_6+t_1) t_2+s_1 s_2
(s_4-t_1+t_2-t_3)-s_5^2 (s_6+2
t_1+t_2-t_3)\ri.}\cr
&\sss{\lf.-(s_2-t_1) t_2 t_3+s_4 (s_5-t_2) (s_5-s_6-t_1+t_3)+s_5
[s_6 (t_1+t_2)+t_1 \
(t_1+2 t_2)-(t_1+t_2) t_3+s_2 (-s_6+t_3)]\ri\} \ F_3}\cr
&\sss{+(s_4+s_5-t_1) [s_1 s_2+(s_5-t_2) \
(-s_3+s_5-t_1+t_3)] \ F_4}\cr
&\sss{+\lf[-s_1^2 s_2+(s_4+s_5-t_1) (s_3-s_5+t_1) (s_5-t_2)
-s_1 (s_4+s_5-t_1) (s_2+s_5-t_2)+
s_1 s_2 t_3\ri] \ F_5}\cr
%%%%%%%%%%%%%%%%%%%%%%%%%%%%%%%%%%%%%%%%%%%%%%%%%%
\sss{s_2\ s_5\ K_4}&=\sss{s_2 s_6 (s_3-t_2) \ F_1+s_6 [s_1 s_2+(s_4+s_5-t_1)
(s_3-t_2)]\ F_2+(s_4+s_5-t_1) (s_3-t_2) (s_1+s_3-s_5-t_3) \ F_6}\cr
&\sss{+\lf\{s_1^2 s_2-s_3^2 \ (s_4+s_5-t_1)+s_3 (s_4+s_5-t_1)
(s_5-s_6-t_1+t_2)+s_3 (s_2+s_4+s_5-t_1) t_3\ri.}\cr
&\sss{\lf.-(s_4+s_5-t_1) t_2 (s_5-s_6-t_1+t_3)-s_1 s_2 (s_3-s_4+t_1-t_2+t_3)-s_2 [s_5
(s_6-t_3)+t_2 t_3)]\ri\}\ F_3}\cr
&\sss{+(s_4+s_5-t_1) [s_1 s_2-(s_3-t_2) (s_3-s_5+t_1-t_3)]\ F_4}\cr
&\sss{+\lf[(s_4+s_5-t_1) (s_3-s_5+t_1) (s_3-t_2)-s_1^2 s_2-s_1 (s_4+s_5-t_1)
(s_2+s_3-t_2)+s_1s_2 t_3\ri]\ F_5}\cr
%%%%%%%%%%%%%%%%%%%%%%%%%%%%%%%%%%%%%%%%%%%%%%%%%%
\sss{s_2\ s_5\ K_{5}}&=\sss{-s_2 s_3 (s_5+s_6) \ F_1-s_3 s_6 (s_4+s_5-t_1) \ F_2
+s_3 (s_4+s_5-t_1) \lf[(s_5+t_3-s_1-s_3)\ F_6+(s_3-s_5+t_1-t_3)\ F_4\ri]}\cr
&\sss{+s_3 [s_1 s_2+(s_4+s_5-t_1) \
(s_3-s_5+s_6+t_1-t_3)-s_2 (s_5+t_3)] \ F_3+s_3 \
(s_1-s_3+s_5-t_1) (s_4+s_5-t_1) \ F_5\ ,}\cr
%%%%%%%%%%%%%%%%%%%%%%%%%%%%%%%%%%%%%%%%%%%%%%%%%%
\sss{s_2\ s_5\ K_{6}}&=\sss{-s_2 (s_5+s_6) (s_3-t_2) \ F_1+
\lf[s_6 (s_4+s_5-t_1) \
(t_2-s_3)-s_1 s_2 s_6+s_2 s_5 (s_6-t_3)\ri]\ F_2}\cr
&\sss{+(s_4+s_5-t_1)(s_3-t_2) (s_5+t_3-s_1-s_3)\ F_6
-(s_4+s_5-t_1) [s_1 s_2-(s_3-t_2) (s_3-s_5+t_1-t_3)] \ F_4}\cr
&\sss{+\lf\{s_1 s_2 (s_3-s_4+t_1-t_2+t_3)-s_1^2 s_2+(s_3-t_2)[(s_4+s_5-t_1)
(s_3-s_5+s_6+t_1-t_3)-s_2 (s_5+t_3)]\ri\}\ F_3}\cr
&\sss{+\lf\{s_1^2 s_2-(s_4+s_5-t_1) (s_3-s_5+t_1) \
(s_3-t_2)+s_1 [(s_4+s_5-t_1) (s_3-t_2)+s_2 (s_4-t_1-t_3)]\ri\}\ F_5\ .}}$$

\appendix\appC{Material for seven--point function}

In this Appendix we present some additional material for the seven--point
function \Becomess, which has been computed in Subsection 3.3.
The complete amplitude is specified by $24$ integrals
$\{K_1,\ldots,K_{24}\}$, whose integrands contain the common factor \Integrandd.
The integrals may be written in the following way
$$K_i=\INTT\ P_i\ \Ic(x,y,z,w)\ ,$$
with the following $24$ polynomials:
\eqn\allsevenF{\eqalign{
P_1&=-\fc{s_{34}}{xy w(1-z)(1-zw)}\ \ ,\ \
P_2=\lf(1-s_{24}+\fc{s_{24}}{x}\ri)\ \fc{1}{yw(1 - z)(1 - zw)(1 -x y z)}\ ,\cr
P_3&=-\fc{s_{34}}{xyzw (1-z)(1-w)}\ \ ,\ \
P_4=\lf(1-s_{24}+\fc{s_{24}}{x}\ri)\ \fc{1}{yzw(1 - z)(1 - w) (1 - x y) }\ ,\cr
P_5&=\fc{s_{34}}{xy(1-z)(1-zw)}\ \ ,\ \
P_6=\lf(\fc{1-s_{14}}{x}+s_{14}\ri)\ \fc{z}{(1 - z) (1 - zw) (1 - x y z)}\ ,\cr
P_7&=\fc{s_{34}}{xy(1-z)(1-w)}\ \ ,\ \
P_8=\lf(\fc{1-s_{14}}{x}+s_{14}\ri)\ \fc{1}{(1 - z)(1 - w) (1 - x y) }\ ,}}
$$\eqalign{
P_9&=-\lf(1-s_{24}+\fc{s_{24}}{x}\ri)\ \fc{1}{y (1 - z) (1 - zw) (1 -xyzw)}\ ,\cr
P_{10}&=-\lf(\fc{1-s_{14}}{x}+s_{14}\ri)\ \fc{z}{(1 - z)
(1 - w z) (1 - x yzw)}\ ,\cr
P_{11}&=-\lf(1-s_{24}+\fc{s_{24}}{x}\ri)\ \fc{1}{y (1 - z) (1 - w) (1 - xyzw)}\ ,\cr
P_{12}&=-\lf(\fc{1-s_{14}}{x}+s_{14}\ri)\ \fc{1}{(1-z)(1-w)(1-xyzw)}\ \ ,\ \
P_{13}=-\fc{s_{34} }{xyz(1-w)(1-zw)}\ ,\cr
P_{14}&= \lf(1-s_{24}+\fc{s_{24}}{x}\ri)\ \fc{1}{y z (1 - w) (1 - x y) (1
-zw)}\ \ ,\ \ P_{15}=\fc{s_{34} }{xy(1-w)(1-zw)}\ ,\cr
P_{16}&= \lf(\fc{1-s_{14}}{x}+s_{14}\ri)\ \fc{1}{(1 - w)
(1 - x y) (1 -zw)}\ ,\cr
P_{17}&=- \lf(1-s_{24}+\fc{s_{24}}{x}\ri)\ \fc{1}{y(1 - w)
(1 - zw) (1 - x yz)}\ ,\cr
P_{18}&=- \lf(\fc{1-s_{14}}{x}+s_{14}\ri)\ \fc{1}{(1 - w)
(1 - zw) (1 - x yz)}\ ,}$$
and:
\eqn\allsevenFF{\hskip-0.5cm\eqalign{
P_{19}&=\lf[\lf(\lf(1-s_{24}+
\fc{s_{24}}{x}\ri)\fc{1}{yz(1-w)}-\lf(\fc{1-s_{14}}{x}+s_{14}\ri)\ri)\fc{1}{1-wz}+
\fc{(1-x)s_{47}}{x(1 -yzw)}\ri]\fc{1}{w(1-z)(1-xy)},\cr
P_{20}&=\lf[\lf(\fc{1-s_{14}}{x}+s_{14}\ri)\fc{1}{(1-zw) (1 -xyz)}-
\fc{s_{34}}{(1-w)xyz}-\fc{(1-x)s_{47}}{x(1 - yzw) (1 -xyz)}\ri]\fc{1}{w(1-z)},\cr
P_{21}&=\lf[\lf(\fc{1-s_{14}}{x}+s_{14}\ri)\fc{1}{(1-w)}-
\lf(1-s_{24}+\fc{s_{24}}{x}\ri)\fc{1}{y(1-zw)}-\fc{(1-x)s_{47}}{x(1 - y zw)}
\ri]\fc{1}{(1-z)(1-xy)},\cr
P_{22}&=\lf[\lf(1-s_{24}+\fc{s_{24}}{x}\ri)\fc{z}{y(1-zw)(1-xyz)}+
\fc{s_{34}}{xy(1-w)}-\fc{z(1-x)s_{47}}{x(1-xyz)(1 - yzw)}
\ri]\fc{1}{(1-z)},\cr
P_{23}&=\lf[\lf(\fc{1-s_{14}}{x}+s_{14}\ri)\fc{1}{(1-w)}
+\lf(s_{34}-\fc{s_{34}}{xy}\ri)\fc{1}{1-zw}-
\fc{(1-x)s_{47}}{x(1 - yzw)}\ri]\fc{(-1)}{(1-z)(1-xyzw)},\cr
P_{24}&=\lf[\lf(1-s_{24}+\fc{s_{24}}{x}\ri)\fc{1}{y(1-w)}-
\lf(s_{34}-\fc{s_{34}}{xy}\ri)\fc{z}{1-zw}-\lf(s_{47}-\fc{s_{47}}{x}\ri)\fc{z}{(1 - yzw)}
\ri]\cr
&\hskip2cm\times\fc{(-1)}{(1-z)(1-xyzw)}\ .}}

\listrefs
\end